\documentclass[useAMS,usenatbib,twocolumn]{mn2e}
\pdfoutput=1
\usepackage{amssymb}
\usepackage{amsmath}
\usepackage{natbib}
\usepackage[pdftex]{graphicx}
\usepackage{subfigure}
\usepackage{booktabs}

\begin{document}

\title[Gravity wave stability analysis]
{Stability analysis of a tidally excited internal gravity wave near the centre of a
  solar-type star}
      \author[A.J. Barker \& G.I. Ogilvie]{Adrian J. Barker\thanks{E-mail:
	  ajb268@cam.ac.uk} and Gordon I. Ogilvie \\
	Department of Applied Mathematics and Theoretical
	Physics, University of Cambridge, Centre for Mathematical Sciences, \\ Wilberforce Road,
	Cambridge CB3 0WA, UK}
	
\date{Accepted 2011 June 23.  Received 2011 June 23; in original form 2011 May 12}

\pagerange{\pageref{firstpage}--\pageref{lastpage}} \pubyear{2011}

\maketitle

\label{firstpage}

\begin{abstract}
We perform a stability analysis of a tidally excited nonlinear internal gravity
wave near the centre of a solar-type star in two-dimensional
cylindrical geometry. The motivation is
to understand the tidal interaction between short-period planets and
their slowly rotating solar-type host stars, which involves the launching of internal gravity
waves at the top of the radiation zone that propagate towards the
centre of the star. Studying the instabilities of
these waves near the centre, where nonlinearities are most important,
is essential, since it may have implications for the survival of short-period
planets orbiting solar-type stars. When these waves have sufficient
amplitude to overturn the
stratification, they break and form a critical layer, which
efficiently absorbs subsequent ingoing wave angular momentum, and can
result in the planet spiralling into the star. However, in previous simulations
the waves have not been
observed to undergo instability for smaller amplitudes. Here we perform a stability
analysis of a nonlinear standing internal gravity
wave in the central regions of a solar-type star. This work has two aims: to determine any
instabilities that set in for small-amplitude waves, and to further understand
the breaking process for large-amplitude waves that overturn the
stratification. Our results are compared with the
stability of a plane internal gravity wave in a uniform stratification, and with previous
work by Kumar \& Goodman on a similar problem to our own. Our main
result is that the waves undergo
parametric instabilities for any amplitude (in the absence of viscosity
and thermal conduction). However, because the nonlinearity
is spatially localised in the innermost wavelengths, the growth rates of these
instabilities tend to be sufficiently small that they do not result in
astrophysically important tidal dissipation. Indeed, we estimate that the
modified tidal quality factors of the star that result are $Q^{\prime}_{\star} \gtrsim
10^{7}$, and possibly much greater, which implies that the resulting
dissipation is at least two orders of magnitude weaker than
that which results from critical-layer absorption. These results support
our explanation for the survival of all currently observed
short-period planets around solar-type main-sequence stars: that
planets unable to cause wave breaking at the centre of their host
stars are likely to survive against tidal decay. This hypothesis will be tested by ongoing and future
observations of transiting planets, such as WASP and Kepler.
\end{abstract}

\begin{keywords}
planetary systems -- stars: rotation --
binaries: close -- hydrodynamics -- waves -- instabilities
\end{keywords}

\section{Introduction}

The tidal interaction between a short-period planet and its host star
can result in evolution of the stellar and planetary spins and the planetary orbit. In particular,
dissipation of the energy stored in the tidal response in the star can
result in the planet spiralling into the star when the period of the
stellar spin is longer than the orbital period. This is because a
final state in which the star spins synchronously with the planetary orbit cannot be achieved when the angular momentum of the orbit is at most
comparable with that of the stellar spin (\citealt{Counselman1973}; \citealt{Hut1980}). In addition, the star
constantly loses spin angular momentum as a result of magnetic
braking, which means that a synchronous equilibrium state does not
exist, and the resulting tidal torque eventually acts to pull the planet towards
the star \citep{Barker2009}. The efficiency of tidal evolutionary processes
depend on the dissipative properties of the star and planet, which are usually parametrised
by a dimensionless quality factor\footnote{Related to the
  traditional $Q$ by $Q^{\prime} =
3Q/2k$, where $k$ is the second-order potential Love number
of the body.} $Q^{\prime}$ for each body, which is an inverse
measure of the dissipation. This is usually defined to be
proportional to the ratio of the maximum energy stored in a tidal oscillation to the energy dissipated
over one cycle (e.g.~\citealt{GoldSot1966}). The mechanisms that
contribute to $Q^{\prime}$ for fluid bodies are poorly understood, but
it is thought that $Q^{\prime}$ depends on tidal
frequency, the internal structure of the body, and, in some cases, the amplitude of
the tidal forcing. In this paper, we study the mechanisms
of tidal dissipation in solar-type main-sequence stars, continuing an
investigation described in previous work by the authors (\citealt{Gio2007},
hereafter OL07; \citealt{Barker2010}, hereafter BO10; \citealt{Barker2011},
hereafter B11).

The response of a fluid body to tidal forcing can be decomposed into
a prolate spheroidal quasi-hydrostatic bulge, referred to as the
equilibrium tide, and a residual
wave-like response, which results from the nonzero forcing frequency
in the frame of the fluid, often referred to as the dynamical tide. In radiation zones of solar-type
stars, the dynamical tide takes the form of internal (inertia) gravity waves (IGWs), which
propagate at frequencies below the buoyancy frequency $N$. These have
previously been proposed to contribute to $Q^{\prime}$ for solar-type
stars (e.g.~\citealt{GoodmanDickson1998}, hereafter GD98; \citealt{Terquem1998}). 

A short-period planet efficiently excites IGWs at the top of the
radiation zone (hereafter RZ),
where their exists a location at which $N\sim 1/P$, with $P$ being the planetary orbital period.
These waves propagate downwards into the RZ, until they
reach the centre of the star, where they are geometrically focused
and can become nonlinear. If their amplitudes are sufficiently large
for the wave to overturn the isentropes\footnote{In stars, the
  stratification is actually composed of both entropy and composition
  gradients. When we refer to ``isentropes'' we actually mean
  stratification surfaces, however, these are usually approximately
  the same.}, the wave breaks and deposits
its angular momentum to form a critical layer, at which ingoing waves
are efficiently absorbed. This results in a strong tidal torque, which
can prevent the survival of sufficiently massive short-period
planets around solar-type stars. However, it only occurs if the planet is sufficiently massive,
or the centre of the star is sufficiently stably stratified. None of
the planets currently observed to orbit solar-type main-sequence stars
is likely to excite waves that break, which could be an
important explanation for their survival.

In BO10 and B11 we performed two- and three-dimensional simulations
of these waves as they approach the centre of the star. The results
were found to be very similar in both two and three dimensions, and if the
amplitude is insufficient for the waves to overturn the
isentropes, the waves were observed to reflect approximately perfectly
from the centre of the star, and no instability appeared to set
in. However, this could be a result of insufficient spatial
resolution or run-time in the simulations performed thus far. In this
paper we perform a detailed stability analysis of a standing internal
gravity wave in two dimensions. The aim of this work is to determine
whether any instabilities are 
likely to occur in reality for
small-amplitude waves, which are unable to overturn the isentropes. 
If an instability exists for these waves, and if this results in efficient
tidal dissipation, then this could have important consequences for the
survival of short-period planets around solar-type stars. 

\subsection{Stability analyses of IGWs}

Many stability analyses of a plane IGW in Cartesian geometry with a uniform
stratification have been performed
(e.g.~\citealt{McEwanRobinson1975}; \citealt{Meid1976};
\citealt{Drazin1977}; \citealt{Klostermeyer1982}). These indicate that
a monochromatic propagating plane IGW is always unstable to parametric
instabilities, whatever its amplitude, in the absence of viscosity and
thermal (or compositional) diffusion. In
that problem such analyses were made possible for finite-amplitude (in
addition to infinitesimal amplitude) waves
because the solution is exact. This is a consequence of the fact
that the wavevector $\mathbf{k}$ and velocity
$\mathbf{u}$ satisfies $\mathbf{k} \cdot \mathbf{u} = 0$, implying
that the advective operator $\mathbf{u} \cdot \nabla$ annihilates any
disturbance belonging to the same plane wave. These stability analyses
allow a detailed understanding of the initial stages of the breaking
process for these waves (e.g.~\citealt{Drazin1977};
\citealt{Klostermeyer1982}; \citealt{LombardRiley1996}; or the review: \citealt{Staquet2002}). 

When a small perturbation is added to a basic plane wave, the resulting
evolutionary equations have periodic coefficients. This allows the
possibility for parametric
instability to occur. The first study of this problem was by
\cite{McEwanRobinson1975}, 
who considered perturbations with length scales much
smaller than the primary IGW wavelength, in which case the problem can
be reduced to the solution of Mathieu's equation. The motion of the
fluid in the basic wave gives rise to unstable modes, just as
parametric oscillations of a pendulum are excited by periodic changes
of its length. The growth rates of these parametrically unstable modes
increase (linearly) with the amplitude of the basic wave.

Subsequent analyses expanded the perturbation onto a Floquet basis,
and relaxed the small-scale assumption. These studies all found that, in
a dissipationless fluid, the disturbances with the largest growth rates
have the smallest spatial scales (e.g.~\citealt{Drazin1977};
\citealt{Klostermeyer1982}). In viscous or radiative fluids, dissipative effects scale with the inverse square
of the length scale of a given mode. This means that the most unstable wavelengths
will no longer be those of the smallest spatial scale, but will be
those for which the competing effects of dissipation and
(nonlinear) growth favour the latter, and this will depend on the Reynolds
number (also the Prandtl number when thermal diffusion is included).

\cite{LombardRiley1996} \& \cite{SonmorKlaassen1997} 
performed a detailed stability analysis of a plane IGW, both
demonstrating that the instability that contributes to wave breaking
is driven by a combination of wave shear and wave entropy
gradients. They find that wave-wave resonance interactions are the
primary mode of instability for small-amplitude waves, with the
picture being much more complicated near overturning
amplitudes. However, no difference
in the source of free energy driving the instability is found for waves that do and do not overturn the
stratification for some wave phase. Some of their results have been
confirmed in recent high resolution numerical simulations (e.g.~\citealt{Fritts2009}).
We discuss these stability analyses further in
relation to our results in \S \ref{comparisonIGW}.

\subsection{This work}

In our problem we have obtained an exact
2D standing wave solution in cylindrical geometry representing IGWs near the centre of a
solar-type star. This enables us to perform a stability
analysis of this wave for any amplitude. This is the subject of the present
paper. One important difference between our problem and
previous studies is that the nonlinearity is
spatially localised to the innermost wavelengths, whereas for
  the plane IGW problem, though the nonlinearity may be localised
  within each wavelength, there is periodic repitition in space.

In the centre of a star, (molecular) viscous damping is negligible,
and the dominant linear dissipation mechanism is radiative
diffusion. However, the waves excited by planets orbiting solar-type stars with several-day
periods have much larger frequencies than their radiative damping
rates. This means that parametrically excited modes with scales
shorter than the primary wave could be produced. These will be
damped by diffusion themselves, but not before they can draw energy from the primary
wave, and possibly contribute to wave breaking.

We have already demonstrated through direct numerical simulations
(BO10; B11) that a wave with sufficient amplitude to
overturn the stratification undergoes a rapid instability (with a
growth time on the order of a wave period) which leads
to wave breaking. We found that the wave
overturns the stratification during part of its cycle if the
angular velocity in the wave exceeds the angular pattern speed of the
forcing, i.e.~$u_{\phi}/r \gtrsim \omega/m$, where $\omega$ is the
wave frequency and $m$ is its azimuthal wavenumber. This can be
expressed as the following breaking criterion derived in B11 valid
for the current Sun. The tidally excited waves break near the centre
of the star if the dimensionless nonlinearity in the wave
\begin{eqnarray}
\label{breakingcriterion}
A \approx 0.28\left(\frac{C}{C_{\odot}}\right)^{\frac{5}{2}}\left(\frac{m_{p}}{M_{J}}\right)\left(\frac{M_{\odot}}{m_{\star}}\right)\left(\frac{P}{1 
\;\mathrm{day}}\right)^{\frac{1}{10}} \gtrsim 1,
\end{eqnarray}
where $m_{\star/p}$ is the mass of the star/planet,
and $C$ is defined such that $N=Cr$ near the centre of the star. The
parameter $A$ is defined so that the wave overturns the isentropes
for some location in the wave if $A\geq1$ (this is defined consistently
with Eq.~\ref{primarywave} below). This means that a one-day
Jupiter-mass planet is not likely to excite IGWs with sufficient
amplitudes to cause breaking at the centre of the
current Sun. However, there is a strong dependence on $C$, which
measures the strength of the stratification near the centre of the
star, which implies that breaking is more likely in older and more
massive stars (of solar-type, with radiative cores).

In the 2D simulations,
the wave reflects perfectly from the centre of the star if its
amplitude is insufficient to satisfy the breaking criterion, and long-term
integrations do not show that any instabilities act on the waves. 
The picture in 3D is very similar. In this
paper we perform a weakly nonlinear stability analysis of our 2D wave
solution using a Galerkin spectral method. This work has
two main aims: 
\begin{itemize}
\item
to better understand the early stages of the breaking process for large-amplitude waves ($A\geq 1$);
\item to
determine what (if any) instabilities may set in for
waves that are unable to overturn the isentropes at any location in
the wave ($A< 1$).
\end{itemize} 

The motivation for this study is that if the waves are subject to
parametric instabilities (as proposed by \citealt{GoodmanDickson1998},
hereafter GD98; \citealt{Kumar1996}; hereafter KG96), whatever their
amplitudes, the reflection of waves from the centre of the star will
not be perfect. This would stand in contrast to the prediction from
linear theory, and the results of our numerical simulations. The
simulations performed thus far may not have the spatial resolution or
have long enough run time to be able to capture small-scale parametric
instabilities. Alternatively, the adopted boundary conditions may
exclude the existence of parametric instabilities. If they indeed
occur in reality, and the tidally excited waves are weakly nonlinearly damped by
parametric instabilities, this could contribute to the tidal
dissipation, and have implications for the survival of short-period
planets with insufficient masses to satisfy
Eq.~\ref{breakingcriterion} and cause breaking.

The paper is structured as follows. First, we present the
Boussinesq-type model derived in BO10, and obtain an exact wave
solution that represents a standing IGW that is confined within a
circular domain. We then derive the equations governing linear perturbations
to this wave, and expand these perturbations using an appropriate,
complete, set of basis functions. The resulting eigenvalue problem is
solved both for cases in which the wave does and does not overturn the
isentropes, and the properties of the resulting unstable modes are studied. We
particularly concentrate on determining the growth rates of the
instabilities and how these vary with the parameters of the problem,
as well as understanding what is the source of free energy driving
them. This is then followed by a discussion, in which
we compare our results with previous studies of the stability of a
plane IGW. We also compare our results with previous work by KG96 on parametric instabilties of
tidally excited waves, and discuss the implications of our results for
the tidal dissipation in solar-type stars, and to the survival of
short-period planets in orbit around them.

\section{Internal gravity wave stability analysis}

We start with the adiabatic Boussinesq-type system (BO10)
\begin{eqnarray}
\label{MainEqs1}
&&D \mathbf{u} = -\nabla q + \mathbf{r}b, \\
&&D b + C^{2} \mathbf{r}\cdot\mathbf{u}  = 0, \\
\label{MainEqs2}
&&\nabla \cdot \mathbf{u} = 0, \\
\label{MainEqs3}
&&D = \partial_{t} + \mathbf{u}\cdot \nabla,
\end{eqnarray}
where $\mathbf{u}$ is the fluid velocity, $b$ is a buoyancy variable
(proportional to the entropy perturbation) and $q$ is a modified
pressure variable. These equations were derived in BO10 from the
equations of gas dynamics and are able to describe the dynamics of
nonlinear IGWs near the centre of a star where the density is nearly
uniform and the buoyancy frequency is proportional to radius.
In this model $N=C r$, where $C$ is a
constant that measures the strength of the stable stratification at
the centre. This model is valid in the innermost $\lesssim 3 \%$ of a solar-type
star, which contains multiple IGW wavelengths for the waves excited by
short-period planets. Acoustic waves have been filtered out from this
model. We have also omitted viscosity and
thermal conduction from these equations, although we will add these
effects later.

Since we restrict our problem to two dimensions, we can express the velocity field in terms of a streamfunction
$\psi$, defined in polar coordinates $(r,\phi)$ by
\begin{eqnarray}
u_{r} &=& \frac{1}{r}\partial_{\phi}\psi, \\ 
u_{\phi} &=&-\partial_{r}\psi,
\end{eqnarray} 
which automatically enforces the solenoidality constraint on the velocity.
We consider a circular region with $r\in [0,r_{out}]$, taking
$r_{out}=1$, which is an impermeable outer boundary at constant entropy, i.e., $\psi(1,\phi,t) =
b(1,\phi,t) = 0$, to confine the modes. We also adopt an inner
regularity condition at $r=0$, which chooses the
regular solutions of the system\footnote{However, in the computation of the
table of integrals described below, this is replaced by an impermeable
inner boundary at $r_{in} = 10^{-4}$, to avoid the coordinate singularity at the
origin.}. This choice of boundary conditions ensures that the total energy of the perturbations is
conserved, since the energy flux through the boundaries is always
zero (because $u_{r}=b=0$ at $r=0,1$).
We use dimensionless units such that the unit of length
$[L] = r_{out}$, the unit of time $[T] =
N^{-1}_{out} = C^{-1} r_{out}^{-1}$, and hence $C = 1$ in these
units. 

To eliminate the modified pressure variable $q$, we take the curl of
the momentum equation:
\begin{eqnarray}
\partial_{t} (\nabla \times \mathbf{u}) = \nabla \times \mathbf{r}b -
\nabla \times (\mathbf{u}\cdot \nabla \mathbf{u}).
\end{eqnarray}
The $z$-component of this equation gives the vorticity equation, which, together with the buoyancy equation, is
\begin{eqnarray}
\label{system1}
\partial_{t} \zeta + \partial_{\phi}b \hspace{-0.1cm}&=&\hspace{-0.1cm}
 J(\psi,\zeta) \\
\label{system2}
\partial_{t} b + \partial_{\phi}\psi \hspace{-0.1cm}&=&\hspace{-0.1cm}
 J(\psi,b),
\end{eqnarray}
where the vorticity is $\zeta =
-\nabla^{2}\psi$. The nonlinear terms have been written in the form of Jacobians,
defined by
\begin{eqnarray}
J(A,B) = \frac{1}{r}\frac{\partial (A,B)}{\partial (r,\phi)} =
\frac{1}{r}\left[(\partial_{r}A)(\partial_{\phi}B) - (\partial_{\phi}A)(\partial_{r}B)\right].
\end{eqnarray}

We consider a stationary, stably stratified
background containing a nonlinear wave (denoted by subscript $w$), subject to a perturbation (denoted
by primes). That is, we expand
\begin{eqnarray}
b &=& b_{w} + b^{\prime}, \\
\psi &=& \psi_{w} + \psi^{\prime}.
\end{eqnarray}
The linearisation of Eqs.~\ref{system1} and \ref{system2} in terms of
the perturbation is
\begin{eqnarray}
\label{wnsystem1}
\partial_{t} (-\nabla^{2} \psi^{\prime}) + \partial_{\phi}b^{\prime} \hspace{-0.1cm}&=&\hspace{-0.1cm}
 J(\psi_{w},-\nabla^{2}\psi^{\prime}) + J(\psi^{\prime},-\nabla^{2}\psi_{w}) \\
\label{wnsystem2}
\partial_{t} b^{\prime} + \partial_{\phi}\psi^{\prime} \hspace{-0.1cm}&=&\hspace{-0.1cm}
 J(\psi_{w},b^{\prime}) + J(\psi^{\prime},b_{w}),
\end{eqnarray}
which is two equations for two unknowns
$(\psi^{\prime},b^{\prime})$. The nonlinearities in this system provide
coupling between different waves. We neglect the terms
$J(\psi^{\prime},-\nabla^{2}\psi^{\prime})$ and
$J(\psi^{\prime},b^{\prime})$, which is consistent with our weakly nonlinear approach.

\subsection{Exact primary wave solution in 2D}

\begin{figure}
  \begin{center}
    \subfigure{\includegraphics[width=0.45\textwidth]{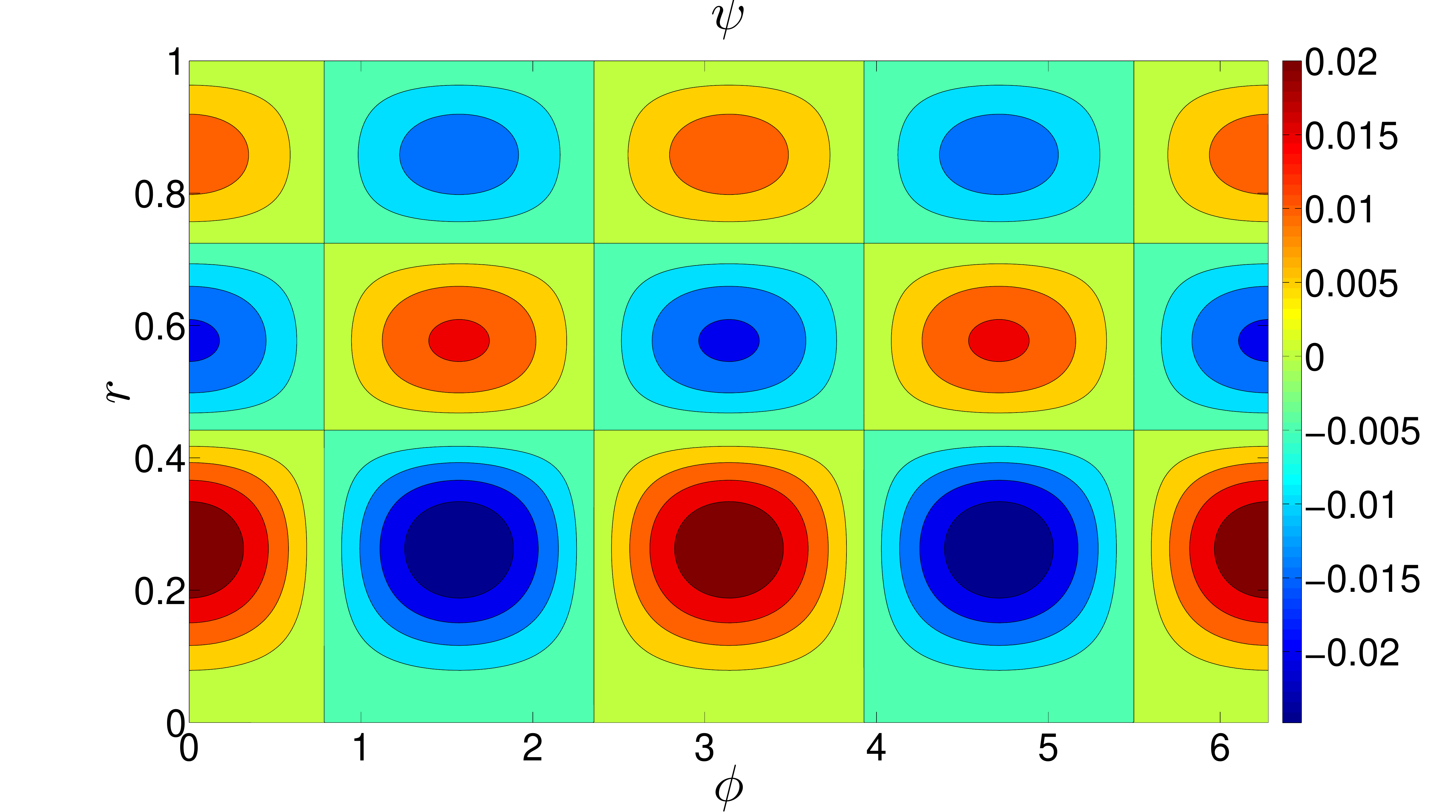} } \\ 
  \end{center}
  \caption{Streamfunction of the primary wave $\psi_{w}$ for $n_{p}=2$, with an arbitrary amplitude for illustration. Contours of
    constant $\psi_{w}$ are the streamlines of the primary wave flow. The flow goes clockwise
    around the red streamlines and anticlockwise around the blue
    streamlines. Stagnation points are located at the radial nodes $r$
    where $J_{2}(k r) = 0$, at azimuthal locations $\phi =
    \frac{(2n+1)}{4}\pi$ for $n\in \mathbb{Z}$.} 
  \label{psicontours}
\end{figure}

We consider a nonlinear gravity wave (the primary wave) with a
well-defined angular pattern speed and azimuthal wavenumber $m=2$. Eqs.~\ref{system1} and
\ref{system2} are invariant under transformation to a rotating frame,
if the streamfunction is transformed appropriately. 
This is because the Coriolis force can be written as the gradient of a
potential, and therefore has no effect.
In the frame in which the wave is steady and $\phi$ is the azimuthal
coordinate, our primary wave is
\begin{eqnarray}
\label{primarywave}
\nonumber
\psi_{w} \hspace{-0.1cm}&=&\hspace{-0.1cm} \;\mathrm{Re}\left[\frac{4}{k^{3}}
  AJ_{2}(k r) e^{2i\phi}\right] 
 \\ \hspace{-0.1cm}&=&\hspace{-0.1cm}
 \frac{2}{k^{3}}\left[AJ_{2}(k r) e^{2i\phi} +
   A^{*}J_{2}(k r) e^{-2i\phi}\right], \\
b_{w}\hspace{-0.1cm} &=&\hspace{-0.1cm} k \psi_{w},
\end{eqnarray}
which is an $m=2$ wave with $n_{p}$ radial nodes (to be chosen
later), where $J_{2}$ is a Bessel function. In general, $A\in
\mathbb{C}$, but is time-independent in this
frame. From here on, we take $A \in \mathbb{R}$, without loss of
generality. Note that $\nabla^{2} \psi_{w} = -k^{2} \psi_{w}$, which
implies that this solution is an exact (nonlinear) solution of Eqs.~\ref{system1}
and \ref{system2}. We choose $k$ such that $J_{2}(k) =
0$. This is equivalent to confining the primary wave in a circular
region of unit radius with
an impermeable outer boundary at constant entropy. We plot an example
of this wave with $n_{p}=2$ in Fig.~\ref{psicontours}.

\begin{figure}
  \begin{center}
    \subfigure{\includegraphics[width=0.45\textwidth]{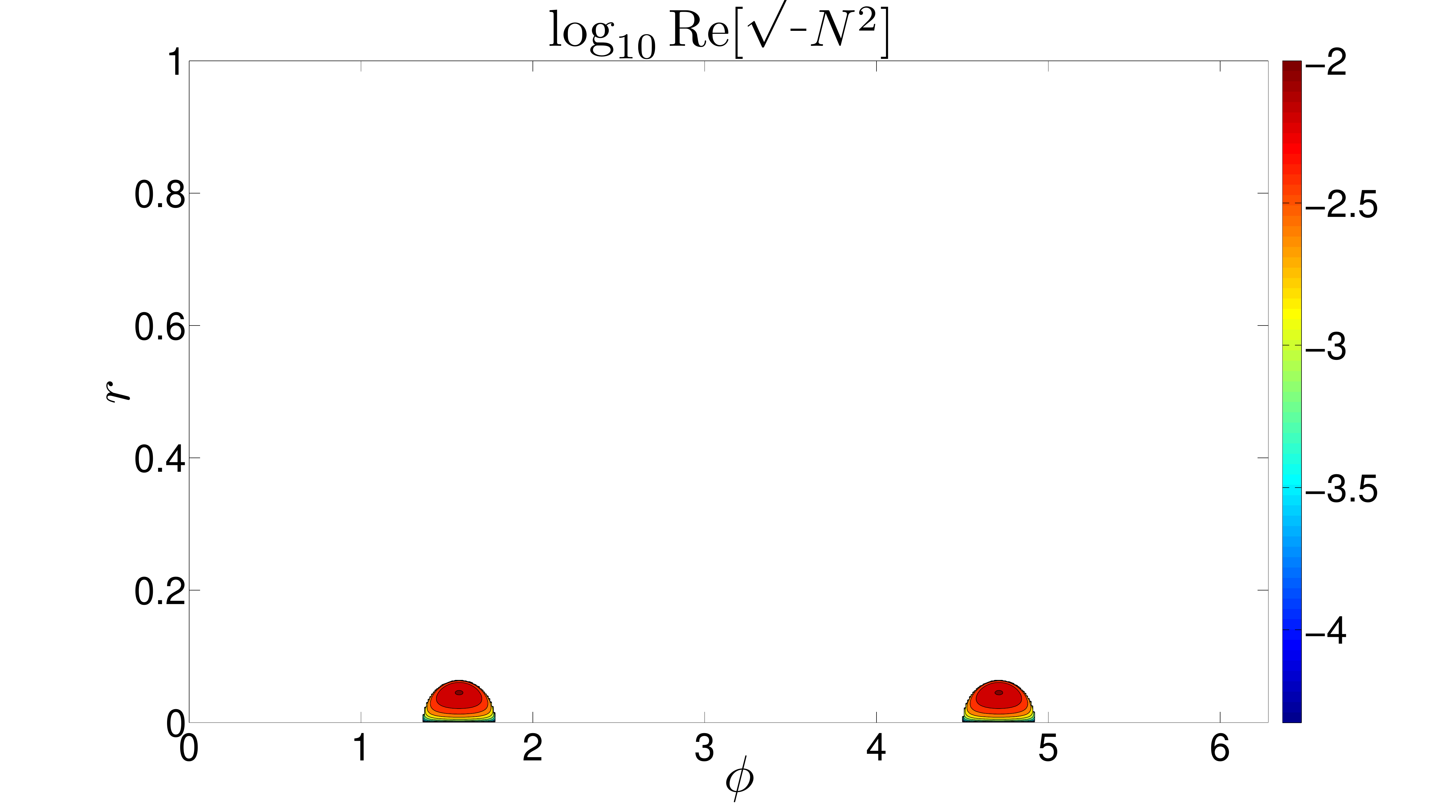} } \\
    \subfigure{\includegraphics[width=0.45\textwidth]{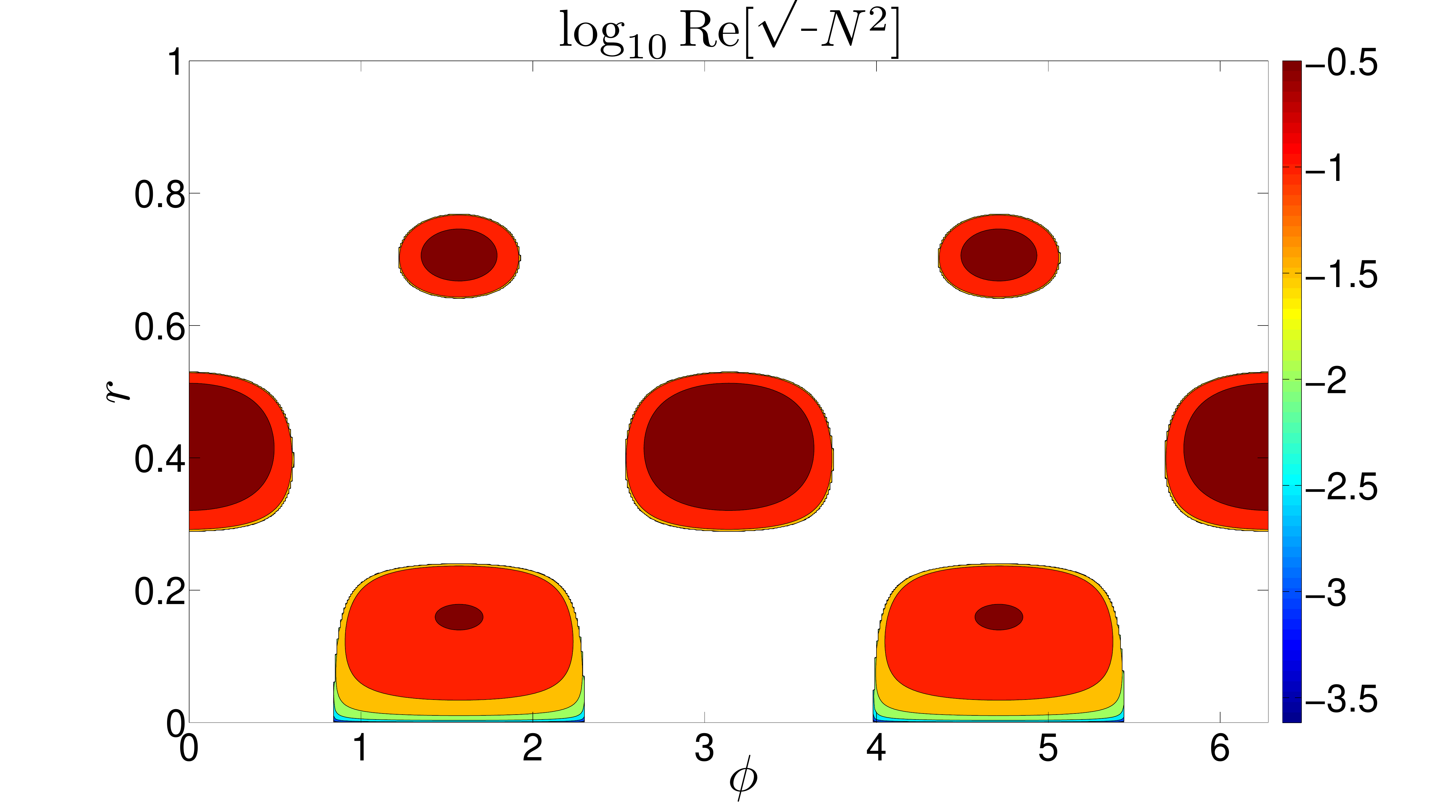} }
  \end{center}
  \caption{Spatial exent of the region that is made convectively unstable by the primary
    wave for $A=1.1$ (top) and $10$ (bottom), with $n_{p}=2$. This region expands from the point $r=0$ when $A=1$ to encompass
    the innermost few wavelengths for larger $A$.}
  \label{N2regionvarA}
\end{figure}

This wave overturns the stratification when $\partial_{r} s < 0$, where $s= (1/2)r^{2} +
b_{w}$ is proportional to the total entropy. This is equivalent to $\frac{1}{r}\partial_{r}b_{w} < -1$.
Note that overturning occurs only when $A>1$, and is more likely
for waves with large radial node numbers and small azimuthal wavenumbers.
The size of the convectively unstable region of the $m=2$ primary wave can be illustrated for a
given $A$ and $n_{p}$, by calculating
\begin{eqnarray}
\nonumber
N^{2} \hspace{-0.1cm} &=& \hspace{-0.1cm} g \partial_{r} s = r(r +
\partial_{r} b) \\
\nonumber
\hspace{-0.1cm} &=& \hspace{-0.1cm} r^{2} + \frac{2
  A}{k}r\left(J_{1}(k r)-J_{3}(k r)\right)\cos 2 \phi.
\end{eqnarray}
For illustration, we plot the
2D region that is convectively unstable for several $A$ values when
$n_{p}=2$ in Fig.~\ref{N2regionvarA}. 
An approximate size for the
overturning region for small $r$ when $A>1$ is
\begin{eqnarray}
r_{ov} \approx \sqrt{\frac{6}{k^{2}}\left(1-\frac{1}{A}\right)},
\end{eqnarray}
at the most unstable wave phase. When $A=1$ the overturning region is the point $r=0$, with the
region expanding for larger $A$. If the instabilities that cause wave
breaking are convectively driven, we would expect them to be strongly
(though not necessarily completely) localised within these
convectively unstable regions of the primary wave.

\subsection{Infinitesimal perturbations}

We consider linear perturbations to this finite-amplitude primary wave,
which we expand as (dropping the primes from now on)
\begin{eqnarray}
\label{galerkinexpansions1}
\psi &=& \sum_{m=-\infty}^{\infty} \sum_{n=0}^{\infty} \psi_{m,n}(t)
J_{m}(k_{m,n} r) e^{i m \phi}, \\
\label{galerkinexpansions2}
b &=& \sum_{m=-\infty}^{\infty} \sum_{n=0}^{\infty} b_{m,n}(t)
J_{m}(k_{m,n} r) e^{i m \phi},
\end{eqnarray}
where $k_{m,n}$ is chosen such that the solutions for each azimuthal
wavenumber $m$, and radial node number $n\geq 0$, satisfy the outer boundary condition at $r=1$, for which
\begin{eqnarray}
J_{m}(k_{m,n}) = 0.
\end{eqnarray}
This condition forces $k_{m,n}\in \mathbb{R}$, for $|m|,n \in
\mathbb{Z}^{+}$. The above expansion automatically enforces a
regularity condition on the perturbations at
$r=0$. Eqs.~\ref{galerkinexpansions1} and \ref{galerkinexpansions2}
define our Galerkin basis. This basis is adopted for two reasons: the
linear solutions ($A=0$) take the same form, and it automatically
satisfies the chosen boundary conditions.

Note that our spectral-space amplitudes $\psi_{m,n}(t),b_{m,n}(t) \in \mathbb{C}$, so we must take
the real part at 
the end of the
calculation to obtain physical quantities. For each $m$, there is an
infinite number of 
components with different
values of $n$. In our spectral representation of the solution, we
truncate these infinite series such that  
$1-L_{m} \leq m \leq L_{m}-1$, where $L_{m}$ is an odd number, and
$0 \leq n \leq L_{n}$. This truncation is chosen so that we have an exactly equal
number either side of $m=0$ (and a similar number either
side of the primary wave $m=2$) which ensures that our mathematical realisation of the problem has the
symmetry property that we discuss in \S~\ref{symmetry} below.

\subsection{Derivation of the evolutionary equations}

Evolutionary equations for the
amplitudes $\psi_{m,n}(t)$ and $b_{m,n}(t)$ can be derived by
projection through integration onto the Galerkin basis. To do this we substitute
the above expansions into the linear system defined by
Eqs.~\ref{wnsystem1}--\ref{wnsystem2}. 
An important orthogonality relation is
\begin{eqnarray}
\label{identity}
\nonumber
&& \hspace{-0.5cm} \int_{0}^{2\pi}\int_{0}^{1} r J_{m^{\prime}}(k_{m^{\prime},n^{\prime}}r)J_{m}(k_{m,n}
r)e^{i(m-m^{\prime})\phi}d r d\phi \\ && \hspace{2.75cm} = \pi\left[J_{m+1}(k_{m,n})\right]^{2}\delta_{n,n^{\prime}}\delta_{m,m^{\prime}},
\end{eqnarray}
where $\delta$ is the Kronecker delta. Note that this results in different normalisation factors for each $m$ and $n$ wave.
Also note that 
\begin{eqnarray}
\label{nablarelation}
-\nabla^{2} \left(J_{m}(k_{m,n} r)e^{im\phi}\right) = k_{m,n}^{2}J_{m}(k_{m,n} r)e^{im\phi}.
\end{eqnarray}

\subsection{Linear solutions (in the absence of the primary wave)}
\label{linearsolution}

Consider Eqs.~\ref{wnsystem1}--\ref{wnsystem2} with $J(\dots,\dots) = 0$, which is equivalent to having a
hydrostatic background with no primary wave
flow. If we substitute the expansions
Eq.~\ref{galerkinexpansions1}--\ref{galerkinexpansions2} into Eqs.~\ref{wnsystem1}--\ref{wnsystem2}, and
then multiply by
$rJ_{m^{\prime}}(k_{m^{\prime}n^{\prime}}r)e^{-im^{\prime}\phi}$, and
finally integrate over $\phi\in[0,2\pi]$ and $r\in [0,1]$, we obtain
\begin{eqnarray}
\label{linearsystem1} 
k_{m,n}^{2}\dot{\psi}_{m,n} + imb_{m,n} = 0 \\
\label{linearsystem2} 
\dot{b}_{m,n} + im\psi_{m,n} = 0
\end{eqnarray}
for each $m,n$, after relabelling $m^{\prime}\rightarrow m$, and
$n^{\prime}\rightarrow n$ following the integration. This system, together with the boundary conditions, can be solved to give
\begin{eqnarray}
\label{linwavesoln}
\psi_{m,n} = A_{m,n}e^{-i \omega_{m,n}t} + B_{m,n}e^{i \omega_{m,n}t},
\end{eqnarray}
with $A_{m,n},B_{m,n} \in \mathbb{C}$, and $\omega_{m,n} = \pm
m/k_{m,n}$. This is the frequency of a non-interacting wave which is able to exist
in the container in the absence of any primary wave flow. If we do not truncate the Galerkin
basis at some finite values of $L_{n}$ and $L_{m}$, then these modes would be dense in the
frequency interval $(0,1)$, since the maximum buoyancy frequency
$N_{max}=1$. In a frame in which the fluid rotates with angular
velocity $\Omega$, the Doppler-shifted wave frequency is
$\hat{\omega}_{m,n} = \omega_{m,n} - m\Omega$. 
Note that $k_{m,n}$ increases with both $m$ and $n$, but $\omega_{m,n}$ decreases
with $n$ and increases with $m$. When substituting the above solution back into
Eq.~\ref{galerkinexpansions1}--\ref{galerkinexpansions2}, we obtain the linear solutions of the
system, which are Bessel functions of a given order $|m|$ with $n$
nodes in the radial direction. This motivated our choice of Galerkin basis.

\subsection{Nonlinear terms}
\label{nonlinearterms}

We obtain our system of equations from Eqs.~\ref{wnsystem1}--\ref{wnsystem2} through the same approach
as in the previous section, to obtain for each $m$ and $n$,
\begin{eqnarray}
\label{weaklynonlinearsystem1}
\nonumber 
\hspace{-0.1cm}k_{m,n}^{2}\dot{\psi}_{m,n} + imb_{m,n} \hspace{-0.25cm}&=&\hspace{-0.25cm}
\frac{1}{\pi
  \left[J_{m+1}(k_{m,n})\right]^{2}}\int_{0}^{2\pi}\int_{0}^{1} \\ \nonumber  && 
\bigg\{rJ_{m}(k_{m,n}r)e^{-im\phi}\left[ J(\psi_{w},-\nabla^{2}\psi) 
\right. \\ && 
\left.
+ J(\psi,-\nabla^{2} \psi_{w})\right]\bigg\}dr d\phi, \\
\label{weaklynonlinearsystem2} 
\nonumber
\hspace{-0.5cm}\dot{b}_{m,n} + im\psi_{m,n} \hspace{-0.25cm}&=&\hspace{-0.25cm} \frac{1}{\pi
  \left[J_{m+1}(k_{m,n})\right]^{2}}\int_{0}^{2\pi}\int_{0}^{1} \\
\nonumber && \hspace{0cm}
\bigg\{rJ_{m}(k_{m,n}r)e^{-im\phi}\left[ J(\psi_{w},b) \right. \\ && \left. 
+ J(\psi,b_{w})\right]\bigg\}dr d\phi,
\end{eqnarray}
where the Jacobians contain sums over $n^{\prime}$ and
$m^{\prime}$. The sum over $m^{\prime}$ is reduced to a pair of terms
through the $\phi$ integration, using Eq.~\ref{identity}. A set of
coupling integrals of triple products of Bessel functions also results, for which there is a sum of such
terms over $n^{\prime}$, i.e., an $m,n$ wave is coupled through
nonlinear terms to waves with $m\pm 2$ and (in principle) all node numbers $n \in \{0,\dots,\infty\}$. The system reduces to 
\begin{eqnarray}
\label{weaklynonlinearsystema} 
\nonumber
\hspace{-0.25cm} k_{m,n}^{2}\dot{\psi}_{m,n} + imb_{m,n} \hspace{-0.25cm}&=&\hspace{-0.25cm} \sum_{n^{\prime}=0}^{\infty}\bigg\{
i\alpha_{m,n,n^{\prime}}\left(k_{m-2,n^{\prime}}^{2}-k_{2,n_{p}}^{2}\right)\tilde{A}\psi_{m-2,n^{\prime}}
\\ && \hspace{-1cm}+
i\beta_{m,n,n^{\prime}}\left(k_{m+2,n^{\prime}}^{2}-k_{2,n_{p}}^{2}\right)\tilde{A}^{*}\psi_{m+2,n^{\prime}} \bigg\},
\\
\label{weaklynonlinearsystemb}
\nonumber
\hspace{-0.25cm} \dot{b}_{m,n} + im\psi_{m,n} \hspace{-0.25cm}&=&\hspace{-0.25cm} \sum_{n^{\prime}=0}^{\infty}\bigg\{
i\alpha_{m,n,n^{\prime}}\tilde{A}\left(
b_{m-2,n^{\prime}}-k_{2,n_{p}}\psi_{m-2,n^{\prime}} \right)
\\ && \hspace{-1cm}+
i\beta_{m,n,n^{\prime}}\tilde{A}^{*} \left(
b_{m+2,n^{\prime}}-k_{2,n_{p}}\psi_{m+2,n^{\prime}} \right)\bigg\},
\end{eqnarray}
where 
\begin{eqnarray}
\tilde{A} = \frac{2}{k_{2,n_{p}}^{3}}A.
\end{eqnarray}
The coupling coefficients are
\begin{eqnarray}
\hspace{-0.25cm}\alpha_{m,n,n^{\prime}} \hspace{-0.25cm}&=& \hspace{-0.25cm}\underbrace{\frac{2}{
  \left[J_{m+1}(k_{m,n})\right]^{2}}}_{\mathrm{normalisation}}\left((m-2)\mathcal{I}^{1}_{m,n,n^{\prime}} -
2\mathcal{I}^{2}_{m,n,n^{\prime}}\right), \\
\hspace{-0.25cm}\beta_{m,n,n^{\prime}} \hspace{-0.25cm}&=& \hspace{-0.25cm}\underbrace{\frac{2}{
  \left[J_{m+1}(k_{m,n})\right]^{2}}}_{\mathrm{normalisation}}\left((m+2)\mathcal{I}^{3}_{m,n,n^{\prime}} + 2\mathcal{I}^{4}_{m,n,n^{\prime}}\right),
\end{eqnarray}
with the integrals
\begin{eqnarray}
\label{int1}
\nonumber
 \mathcal{I}^{1}_{m,n,n^{\prime}}  \hspace{-0.25cm}&=& \hspace{-0.25cm} \int_{0}^{1} J_{m}(k_{m,
  n}r)\left[\partial_{r}J_{2}(k_{2,n_{p}}r)\right]J_{m-2}(k_{m-2,n^{\prime}}r) dr, \\
\label{int2}
\nonumber
 \mathcal{I}^{2}_{m,n,n^{\prime}} \hspace{-0.25cm} &=& \hspace{-0.25cm} \int_{0}^{1} J_{m}(k_{m,
  n}r)J_{2}(k_{2,n_{p}}r)\left[\partial_{r}J_{m-2}(k_{m-2,n^{\prime}}r)\right] dr, \\
\label{int3}
\nonumber
\mathcal{I}^{3}_{m,n,n^{\prime}}  \hspace{-0.25cm}&=&  \hspace{-0.25cm}\int_{0}^{1} J_{m}(k_{m,
  n}r)\left[\partial_{r}J_{2}(k_{2,n_{p}}r)\right]J_{m+2}(k_{m+2,n^{\prime}}r) dr, \\
\label{int4}
\nonumber
 \mathcal{I}^{4}_{m,n,n^{\prime}}  \hspace{-0.25cm}&=& \hspace{-0.25cm} \int_{0}^{1} J_{m}(k_{m,
  n}r)J_{2}(k_{2,n_{p}}r)\left[\partial_{r}J_{m+2}(k_{m+2,n^{\prime}}r)\right] dr.
\end{eqnarray}
Note that these are related by
\begin{eqnarray}
\mathcal{I}^{3}_{m,n,n^{\prime}} &=&
\mathcal{I}^{1}_{m+2,n^{\prime},n} \;, \\
\mathcal{I}^{4}_{m,n,n^{\prime}} &=&
\mathcal{I}^{2}_{m+2,n^{\prime},n} -
\mathcal{I}^{1}_{m+2,n^{\prime},n} \;,
\end{eqnarray}
where the latter follows from an integration by parts. 
For use in the derivation of the spectral space energy equation in a
subsequent section, we find it convenient to define 
\begin{eqnarray}
\tilde{\alpha}_{m,n,n^{\prime}} = \pi \left[J_{m+1}(k_{m,n})\right]^{2}\alpha_{m,n,n^{\prime}},
\end{eqnarray}
and similarly for $\tilde{\beta}_{m,n,n^{\prime}}$. This is because we then
have the relation
\begin{eqnarray}
\tilde{\beta}_{m,n,n^{\prime}} = \tilde{\alpha}_{m+2,n^{\prime},n}.
\end{eqnarray}

\subsection{Diffusive terms}

In the presence of viscosity and radiative diffusion (or hyperdiffusion)
Eqs.~\ref{wnsystem1} and \ref{wnsystem2} have the additional terms $(-1)^{2+\alpha}\nu
\nabla^{2+2\alpha} \psi$ and $(-1)^{1+\alpha} \kappa \nabla^{2\alpha}
b$, respectively. Here $\alpha$ is chosen to give the standard diffusive operator
($\alpha=1$), or hyperdiffusion ($\alpha=2,3$). In this case we obtain the (linearised) system
\begin{eqnarray}
k_{m,n}^{2}\dot{\psi}_{m,n} + imb_{m,n} = -\nu k_{m,n}^{2+2\alpha} \psi_{m,n}, \\
\dot{b}_{m,n} + im\psi_{m,n} = -\kappa k_{m,n}^{2\alpha} b_{m,n},
\end{eqnarray}
instead of
Eqs.~\ref{linearsystem1}--\ref{linearsystem2}, where
$\nu$ is the kinematic (hyper-) viscosity and $\kappa$ is the thermal
(hyper-) diffusivity, with similar modifications to Eqs.~\ref{weaklynonlinearsystema}--\ref{weaklynonlinearsystemb}. The dispersion
relation is then
\begin{eqnarray}
\left(\omega_{m,n}+i \nu k_{m,n}^{2\alpha} \right)\left(\omega_{m,n}+i
  \kappa k_{m,n}^{2\alpha} \right) = \frac{m^{2}}{k_{m,n}^{2}},
\end{eqnarray}
indicating that the frequencies of the allowed solutions are modified in the
presence of $\nu,\kappa$. The growth rate expected in the presence of weak
diffusion is therefore
$\mathrm{Im}\left[\omega\right] - \frac{1}{2}(\nu+\kappa) k_{m,n}^{2\alpha}$, with
$\mathrm{Im}\left[\omega\right]$ being the appropriate inviscid growth
rate. Since Eqs.~\ref{weaklynonlinearsystema}--\ref{weaklynonlinearsystemb}
allow instability, to obtain
growing modes in the presence of diffusion, $\nu$ and $\kappa$ must be
sufficiently weak so that diffusive terms do not dominate over the
nonlinear terms except for values of $n$ and $m$ close to the resolution limits of
$L_{n}$ and $L_{m}$. Hyperdiffusion with $\alpha=3$ is adopted
since it better restricts the dissipation to the highest wavenumbers.
This enables numerical convergence in the eigenvalue problem discussed
below, but does not significantly perturb the growth rates of the
lower wavenumber eigenmodes with the values of $\nu,\kappa$ that we
adopt. From here on, we also take $\nu=\kappa$.

With the inclusion of diffusive terms we require $1+2\alpha$ additional boundary conditions
at each boundary. These are regularity conditions at the centre, and
at the outer boundary we can consider a variety of idealised boundary
conditions of the form $\nabla^{2\sigma}\psi = 0$ for $\sigma =
0,1,\dots,\alpha$ and $\nabla^{2\sigma}b =0$ for $\sigma =
0,1,\dots,\alpha-1$. These are automatically satisfied by our Galerkin basis
Eq.~\ref{galerkinexpansions1}--\ref{galerkinexpansions2} and the
definition of $k_{m,n}$. With these boundary conditions (which force
$k_{m,n}$ to be real), the Galerkin basis is therefore exact for the
single-wave diffusive problem. Note that this is also true if
hyperdiffusion is adopted, due to property Eq.~\ref{nablarelation}. In
the numerical solution of the eigenvalue problem in the following
sections we include these diffusive terms to achieve numerical
convergence. This is necessary because in the absence of diffusion, the
most unstable modes are found to prefer the smallest spatial scales.

\section{Method of solution}

Our system Eqs.~\ref{weaklynonlinearsystem1}--\ref{weaklynonlinearsystem2} can be
written in the form of a generalised eigenvalue problem of the form
\begin{eqnarray}
\label{eigenvalueproblem}
\boldsymbol{\mathrm{A}}\mathbf{U} = \omega \boldsymbol{\mathrm{B}}\mathbf{U},
\end{eqnarray}
where $\mathbf{U}$ is the column vector whose components are the
quantities $(\psi_{m,n},b_{m,n})$ for
each $m$ and $n$. $\boldsymbol{\mathrm{A}}$ is the block tridiagonal matrix
representing the system. This is done by
seeking normal mode solutions of the form $\psi_{m,n}(t) \propto e^{-i\omega t}$ for each $m$ and $n$, and
  similarly for $b_{m,n} (t)$. $\boldsymbol{\mathrm{B}}$ is the
  diagonal matrix that can be composed as blocks of the form
\begin{eqnarray}
 -i\left( 
\begin{array}{cc}
        k_{m,n}^{2} & 0\\
        0 & 1  \\
      \end{array} \right),
\end{eqnarray}
for each $m$ and $n$. We solve this problem using standard generalised eigenvalue
solver routines, such as ZGGEV in the LAPACK library. This returns the eigenvalues
$\{ \omega \}$, and the spectral space eigenfunctions $\{ \psi_{m,n},b_{m,n}
\}$ corresponding to each eigenvalue. The real space eigenfunctions can be reconstructed from these,
using Eqs.~\ref{galerkinexpansions1}--\ref{galerkinexpansions2}.

We choose $L_{n}=50$ and $L_{m}=27$ for most of the
calculations (though a small number of higher resolution calculations were performed
with $L_{m}$ values up to $45$, which confirm that our results are not dependent on resolution). There is a limit to the
  maximum values of $L_{n}$ and $L_{m}$ that we can reasonably adopt,
  due to the computational cost of choosing large values for each of
  these parameters. One
  reason for this is that the nonlinear terms require the computation
of a large number of integrals, many of which have highly oscillatory
integrands, and require very small relative error tolerances to be
computed accurately (by the method of computation that we will describe in the next subsection).
Another reason is that we compute the eigenvalues and eigenvectors
using a QZ alogorithm, which has a high computational cost $O(10 S^3)$
for an $S\times S$ matrix, where $S= 2L_{m}\left(L_{n}-1\right)$. This
limits the values of $L_{n}$ and $L_{m}$.
With our choice of $\alpha=3$
hyperdiffusion, it has been found that
$10^{-14}\leq \nu \leq 10^{-11}$
is appropriate. This hyperdiffusion is found to give the spurious
eigenvalues, whose eigenfunctions oscillate at the smallest
scales, and are therefore not adequately resolved, a large decay rate, and allows our
growing modes to be adequately converged for the values of $A$ and $n_{p}$
that we consider.

\subsection{Numerical computation of table of integrals}

The tables of integrals defined in \S~\ref{nonlinearterms} are computed for each
value of $m,n,n^{\prime}$ using a 4th/5th order adaptive
step Runge-Kutta integrator. To enable efficient computations, the
Bessel functions are computed simultaneously with the integrals. Note that Bessel's equation
\begin{eqnarray}
\partial_{r}(r\partial_{r} \psi) +
r\left(k_{m,n}^{2}-\frac{m^{2}}{r^{2}}\right)\psi = 0,
\end{eqnarray}
can be rewritten as the coupled set of first order ODEs
\begin{eqnarray}
\frac{d\xi}{dr} &=& r\left(\frac{m^{2}}{r^{2}}-k_{m,n}^{2}\right)\psi, \\
\frac{d\psi}{dr} &=& \frac{\xi}{r}.
\end{eqnarray}
We also need the derivatives of various Bessel functions, so we also integrate
\begin{eqnarray}
\frac{d^{2}\psi}{dr^{2}} = \frac{(m^{2}-r^{2}k_{m,n}^{2})\psi-\xi}{r^{2}},
\end{eqnarray}
to obtain the first derivative of each Bessel
function. To compute the integrals, we integrate the integrands of
$\mathcal{I}^{i}_{m,n,n^{\prime}}$ for $i\in \{1,2,3,4\}$,
and take the value at $r=1$. This involves solving a system of 15 ODEs in total for each
$m,n,n^{\prime}$. Using our chosen resolution of $L_{n}=50$, $L_{m}=27$, this
involves the computation of $L_{m} (L_{n}-1)^2 \sim 10^{5}$ integrals in
total. These are computed
for a given number of radial nodes in the primary wave in the range $0
\leq n_{p} \leq 12$. 
We impose an inner boundary at $r_{in}=10^{-4}$ to avoid
the coordinate singularity at the origin, and use initial conditions
appropriate from considering the asymptotic behaviour of the Bessel
functions. For small $L_{n}$ and $L_{m}$ values the numerical integrals have
been checked to agree with those computed from Mathematica, and for
large $L_{n}$ and $L_{m}$, several integrals containing the highest $n$ and
$m$ values were also checked. We use a
relative error tolerance of $10^{-13}$, which has been found to
compute the most oscillatory integrals (corresponding to the highest
$n$ and $m$ value) accurately (compared with Mathematica) to within at
least $6$ decimal places.

\section{Kinetic and potential energy equations}
\label{spectralenergy}

The kinetic and potential energies can be computed from either the
real-space or spectral-space eigenfunctions. This enables a determination of
the dominant source of free energy driving the instability (e.g.~\citealt{LombardRiley1996}), and
also provides an independent calculation of the growth rate, which can be
used to check our numerical code. We derive an energy equation in
spectral space, and compute the volume-integrated terms using the
numerically computed eigenfunctions, without converting to real
space. This has been found to reduce numerical errors, resulting from
large numerical cancellations in the most oscillatory Bessel
functions, when the energy equations are instead computed in real space. 
In addition, it is simpler to construct the (hyper-) diffusion
terms in spectral space, so that they can be fully taken into account in the energy budget.

We define 
\begin{eqnarray}
K &=& \int_{0}^{2\pi}\int_{0}^{1} \frac{1}{2}|\mathbf{u}|^{2} r dr
d\phi, \\
P &=& \int_{0}^{2\pi}\int_{0}^{1} \frac{1}{2}b^{2} r dr
d\phi, \\
E &=& K + P,
\end{eqnarray}
as the kinetic, potential and total
energy densities of the disturbance, respectively. 
Note that
\begin{eqnarray}
 K  &=&
\frac{1}{2}\sum_{m=-\infty}^{\infty}\sum_{n=0}^{\infty}k_{m,n}^{2}|\psi_{m,n}|^{2}\pi
\left[J_{m+1}(k_{m,n})\right]^{2}, \\
P  &=& \frac{1}{2}\sum_{m=-\infty}^{\infty}\sum_{n=0}^{\infty}|b_{m,n}|^{2}\pi
\left[J_{m+1}(k_{m,n})\right]^{2}, \\
E  &=& K  + P.
\end{eqnarray}
Evolutionary equations for the volume-integrated energy can be obtained
from Eq.~\ref{weaklynonlinearsystema} and
\ref{weaklynonlinearsystemb} together with the (hyper-) diffusion terms. After some rearrangement, these can be written
\begin{eqnarray}
\label{Energyeqs1}
\dot{ K } &=&  \mathcal{N}_{sw}  +  F_{b}  +  F_{\nu} , \\
\label{Energyeqs2}
\dot{ P } &=&  \mathcal{N}_{bw}  -  F_{b}  +  F_{\kappa} , \\
\label{Energyeqs3}
\dot{ E } &=&  \mathcal{N}_{sw}  +  \mathcal{N}_{bw}  +  F_{\nu}  +  F_{\kappa} ,
\end{eqnarray}
where
\begin{eqnarray}
\nonumber
 \mathcal{N}_{sw} \hspace{-0.25cm} &=& \hspace{-0.25cm}
\mathrm{Re}\sum_{m=-\infty}^{\infty}\sum_{n=0}^{\infty}\sum_{n^{\prime}=0}^{\infty}i\tilde{\alpha}_{m,n,n^{\prime}}\left(k_{m-2,n^{\prime}}^{2}-k_{m,n^{\prime}}^{2}\right)
  \\ \nonumber && \hspace{4cm} \tilde{A}\psi_{m,n}^{*}\psi_{m-2,n^{\prime}}
, \\ \nonumber
 \mathcal{N}_{bw} \hspace{-0.25cm} &=& \hspace{-0.25cm}
\mathrm{Re}\sum_{m=-\infty}^{\infty}\sum_{n=0}^{\infty}\sum_{n^{\prime}=0}^{\infty}i\tilde{\alpha}_{m,n,n^{\prime}}\tilde{A}k_{2,n_{p}}
   \\ \nonumber && \hspace{2cm}
\left(b_{m,n}^{*}\psi_{m-2,n^{\prime}}+\psi_{m,n}^{*}b_{m-2,n^{\prime}}\right)
    , \\ \nonumber
 F_{b} \hspace{-0.25cm} &=& \hspace{-0.25cm} \mathrm{Re}
\sum_{m=-\infty}^{\infty}\sum_{n=0}^{\infty}im\psi_{m,n}^{*}b_{m,n}\pi
\left[J_{m+1}(k_{m,n})\right]^{2} , \\ \nonumber
 F_{\nu} \hspace{-0.25cm} &=& \hspace{-0.25cm}
\mathrm{Re}\sum_{m=-\infty}^{\infty}\sum_{n=0}^{\infty}-\nu k_{m,n}^{2+2\alpha}|\psi_{m,n}|^{2}\pi
\left[J_{m+1}(k_{m,n})\right]^{2}, \\ \nonumber
 F_{\kappa} \hspace{-0.25cm} &=& \hspace{-0.25cm}
\mathrm{Re}\sum_{m=-\infty}^{\infty}\sum_{n=0}^{\infty}-\kappa k_{m,n}^{2\alpha}|b_{m,n}|^{2}\pi
\left[J_{m+1}(k_{m,n})\right]^{2}.
\end{eqnarray}
$ \mathcal{N}_{sw}$ represents the production of perturbation
kinetic energy from the primary wave shear. $
\mathcal{N}_{bw}$ represents the production of perturbation
potential energy from the primary wave entropy gradients. Whichever of $ \mathcal{N}_{sw}$ or
$ \mathcal{N}_{bw}$ is dominant tells us whether this instability is driven by
the wave shear or the wave entropy gradients. $ F_{b}$
is the buoyancy flux term, representing conversion between kinetic and potential
energies of the disturbance. Finally, $ F_{\nu}$ and
$ F_{\kappa}$ represent the irreversible loss of kinetic and
potential energies as a result of the (hyper-) diffusion.

After truncation at $|m|=L_{m}$ and $n=n^{\prime}=L_{n}$, each of these terms
are computed from the spectral space eigenfunctions, together 
with the numerically computed table of integrals. The growth rate can
then be computed from
\begin{eqnarray}
\label{growthratepredictions}
\mathrm{Im}[\omega{^{\prime}}] = \frac{\dot{ E }}{2 E } = \frac{\dot{ K }}{2 K } = \frac{\dot{ P }}{2 P }.
\end{eqnarray}
We have checked that each of these equations are accurately satisfied to within at worst a
few percent for each of the unstable modes discussed in this
paper. This provides a check of our analytical derivations and 
numerical calculations, and should convince ourselves that our
results are consistent.

\section{Numerical tests}

In this section we briefly mention several numerical tests which we
have performed to validate our numerical code, in addition to the one
mentioned in the previous section. Following this section, in
\S~\ref{parametricresults} and \ref{breakingresults} we discuss the
results of our stability analysis for waves with $A<1$ and $A>1$, respectively.

\subsection{Linear}

In the absence of nonlinear couplings ($A=0$), we obtain a set of
non-interacting modes with eigenfreqencies $\omega \in
\{\omega_{m,n}\}$, where $\omega_{m,n} =
\pm m/k_{m,n}$ (in the inertial frame), as we predicted in \S~\ref{linearsolution}. In the absence of diffusion, these
have zero growth rate, i.e., Im$\left[\omega\right] =
0$, for all eigenmodes. When hyperdiffusion is
included, the eigenmodes each have a nonzero decay rate determined by the
values of $\nu$ and $\kappa$, which is very
accurately (to more than 10 decimal places) computed from considering only the terms $ F_{\nu}$ and
$ F_{\kappa}$ in the energy equation. The real-space eigenfunctions that
result are what is predicted from linear theory, in that they are
Bessel functions of order $m$ with $n$ nodes in the radial
direction.

\subsection{Weakly nonlinear}

We followed a typical eigenmode as $A$ is gradually increased from
zero, and found that for $|A| \ll 1$, the shift in the eigenfrequency $\mathrm{Re}[\delta \omega] \propto
A^{2}$, as we would expect for modes not undergoing parametric
resonance (e.g.~\citealt{Landau1969}). 
\subsection{Symmetries}
\label{symmetry}

The real-space solutions can be represented in the form
\begin{eqnarray}
\mathrm{Re}\left[\sum_{m}\sum_{n}c_{m,n}(r)e^{i(m\phi-\omega t)}\right].
\end{eqnarray}
This is symmetric under the transformations $c_{m,n}(r)\rightarrow
c_{m.n}^{*}(r)$, $m\rightarrow -m$ and $\omega \rightarrow
-\omega^{*}$. This symmetry should exist for all primary wave
amplitudes, and results from the fact that only the sign of the pattern speed
$\omega/m$ has meaning, and not the sign of the wavenumber or frequency. 
This means that when the eigenvalues are plotted on the
complex frequency plane, they should be symmetric about
$\mathrm{Re}\left[\omega\right] =0$.

\section{Results for waves with $A<1$: parametric instabilities}
\label{parametricresults}

\begin{figure}
  \begin{center}
    \subfigure{\includegraphics[width=0.48\textwidth]{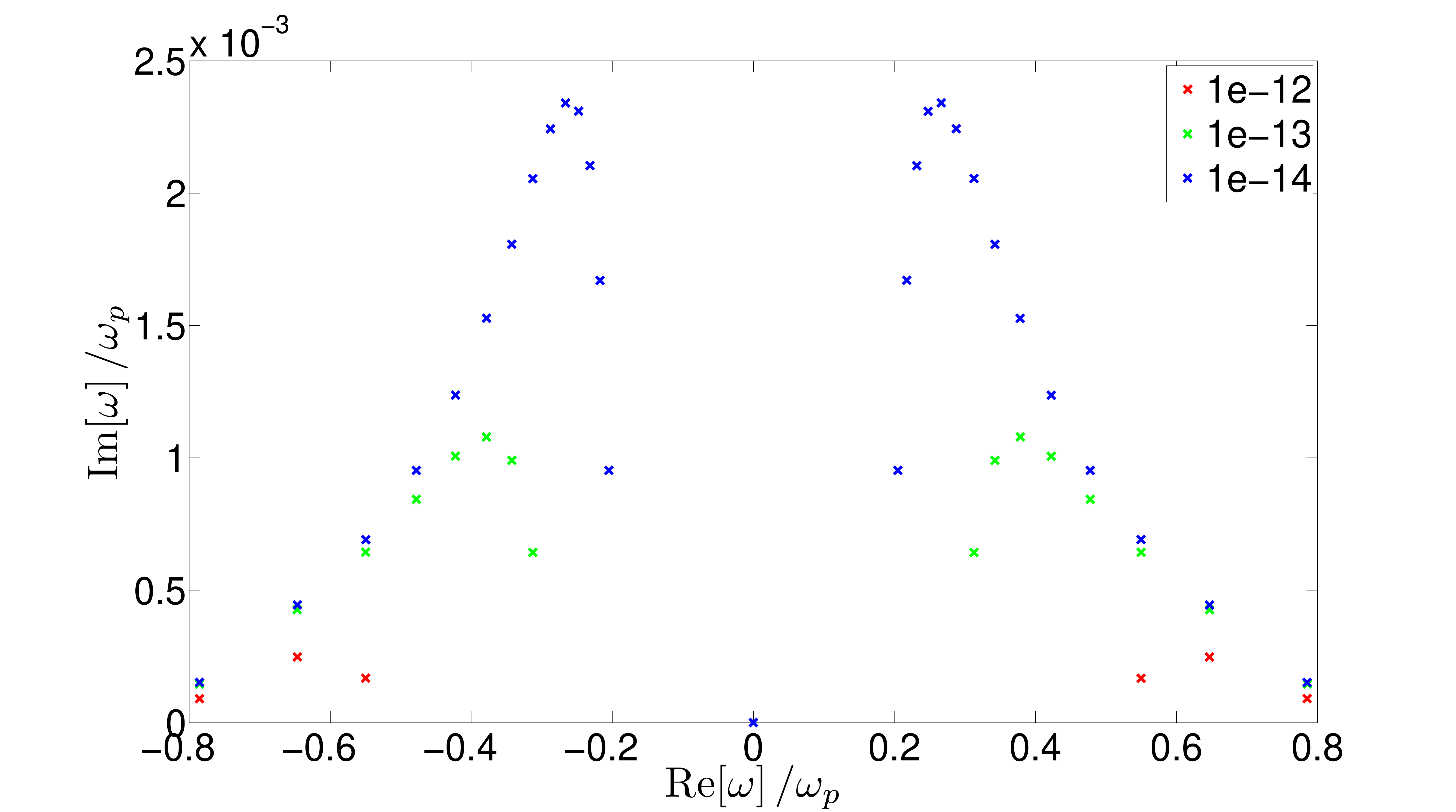} }
\end{center}
  \caption{Distribution of unstable eigenvalues on the complex
    frequency plane for $A=0.1$, $n_{p}=2$ and $\nu=10^{-12},10^{-13},10^{-14}$.} 
  \label{galerkinpar1}
\end{figure}

We examine the unstable modes that exist when $0 < A < 1$, which is
when the primary wave does not overturn the stratification at any
location in the wave. The instability is a parametric
instability, for which a simple model is briefly reviewed in Appendix \ref{parametricinstability}.
When $A\ne 0$, the fraction of eigenmodes
that are growing is nonzero (above a critical $A$ set by the values
of $n_{p}$ and $\nu$, which can be understood from Eq.~\ref{thresholdparametric}) and increases with $A$, as
nonlinear growth starts to dominate over the decay due to diffusion
for a larger number of modes. For small $A$, the eigenvalues of the unstable modes displayed on the
complex plane are distributed in two curves which are symmetric about $\mathrm{Re}\left[\omega\right] =0$.
This is illustrated in Fig.~\ref{galerkinpar1}, and is a result of the symmetry described
in \S~\ref{symmetry}.

In the limit that $A\rightarrow 0$, the unstable modes can be
identified as the parametrically excited free-wave modes. These are a pair of free wave modes that exist
when $A=0$, which undergo modifications to their complex frequencies
at $O(A)$ that reinforce each other. We have verified that the modes
consist of a pair whose frequencies approximately add up to $\omega_{p}$ in the
inertial frame, with a detuning $\Delta/\omega_{p} \lesssim 10^{-2}$. As $A$ is increased, the unstable modes consist of gradually more
complicated superpositions of free wave modes, until for $A \gtrsim 1$, the eigenfunctions become localised in the convectively unstable
regions. This will be studied in more detail in \S~\ref{breakingresults}. For $A\lesssim 1$,
the eigenfunctions exist because of their confinement by the
boundaries, though they interact quite strongly with the primary wave,
and are generally not simply free wave modes with a nonzero growth rate.

\begin{figure}
  \begin{center}
    \subfigure{\includegraphics[width=0.48\textwidth]{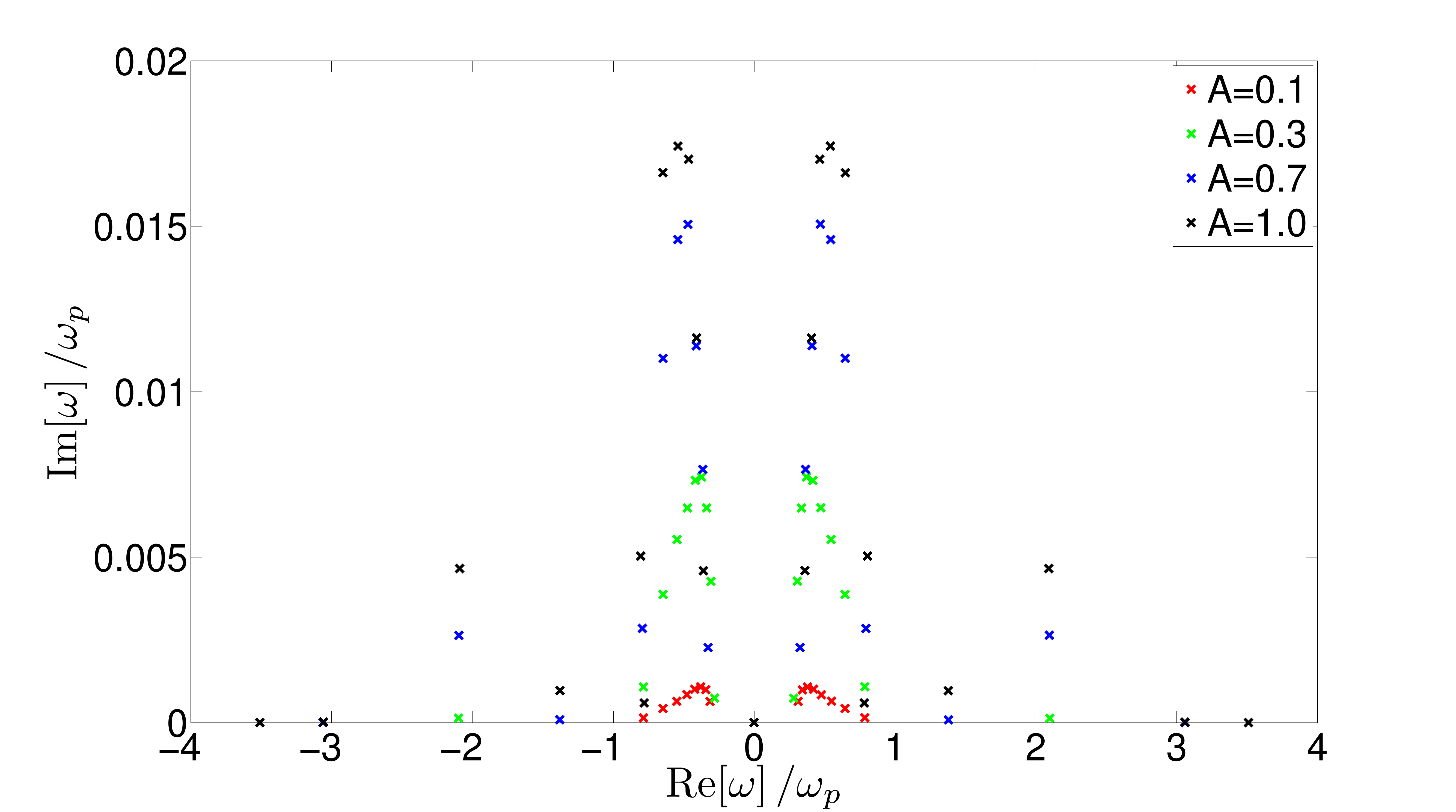} }
\end{center}
  \caption{Distribution of unstable eigenvalues on the complex
    frequency plane for various values of $A$, with $n_{p}=2$ and $\nu=10^{-14}$.} 
  \label{galerkinpar2}
\end{figure}

The number of unstable modes that exist in this amplitude range
depends quite strongly on viscosity. This is illustrated in
Fig.~\ref{galerkinpar1}. The number is also found to decrease as
$n_{p}$ is increased. These two behaviours are related by the fact
the a wave should undergo parametric instabilities, which
have largest growth rates when the resonant tuning is good, which is
more likely to occur for perturbations with larger wavenumbers. However, these
large wavenumber components are strongly damped by diffusion. Increasing
$n_{p}$ means that the ``effective resolution'' available to
capture the unstable modes decreases. This is the same as increasing
$\nu$, hence the same trends exhibited in increasing $n_{p}$ and $\nu$.

The neat distribution of eigenvalues into two curves does not persist
as $A$ is increased, as illustrated in
Fig.~\ref{galerkinpar2}. The frequencies (in the rotating frame) are primarily
smaller than the primary wave frequency in the inertial frame.
However, a small number of modes exist for $A\gtrsim 0.3$ which have
frequencies larger than $\omega_{p}$. Nevertheless, in each case the
frequencies of the unstable modes are always smaller than the maximum
buoyancy frequency in the flow (which corresponds with $1/\omega_{p}
\sim 6$ in the units of this figure, for $n_{p}=2$). This makes sense if these are gravity wave-like disturbances, which are
parametrically excited by the primary wave.

\begin{figure}
  \begin{center}
    \subfigure{\includegraphics[width=0.45\textwidth]{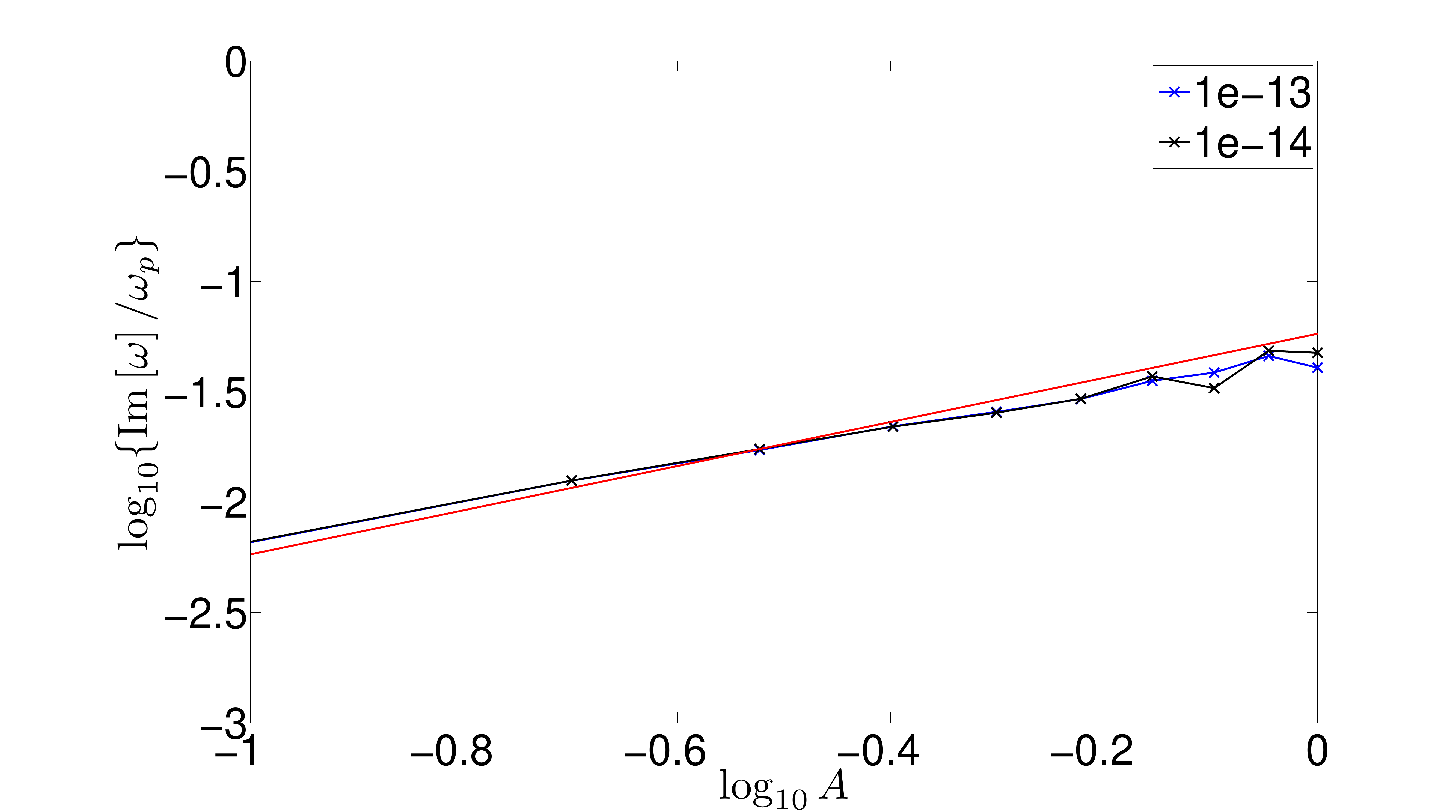} }
  \end{center}
  \caption{Im$\left[\omega\right]$
    vs.~$A$ for the most unstable mode with $n_{p}=2$ and $\nu=10^{-13},10^{-14}$. The solid red line
    has slope $1$. This shows that the instability approximately
    scales linearly with $A$ for small $A$, indicating that the
    instability when $A<1$ is due to a parametric resonance.}
  \label{galerkinpar6}
\end{figure}

\begin{figure}
  \begin{center}
    \subfigure{\includegraphics[width=0.48\textwidth]{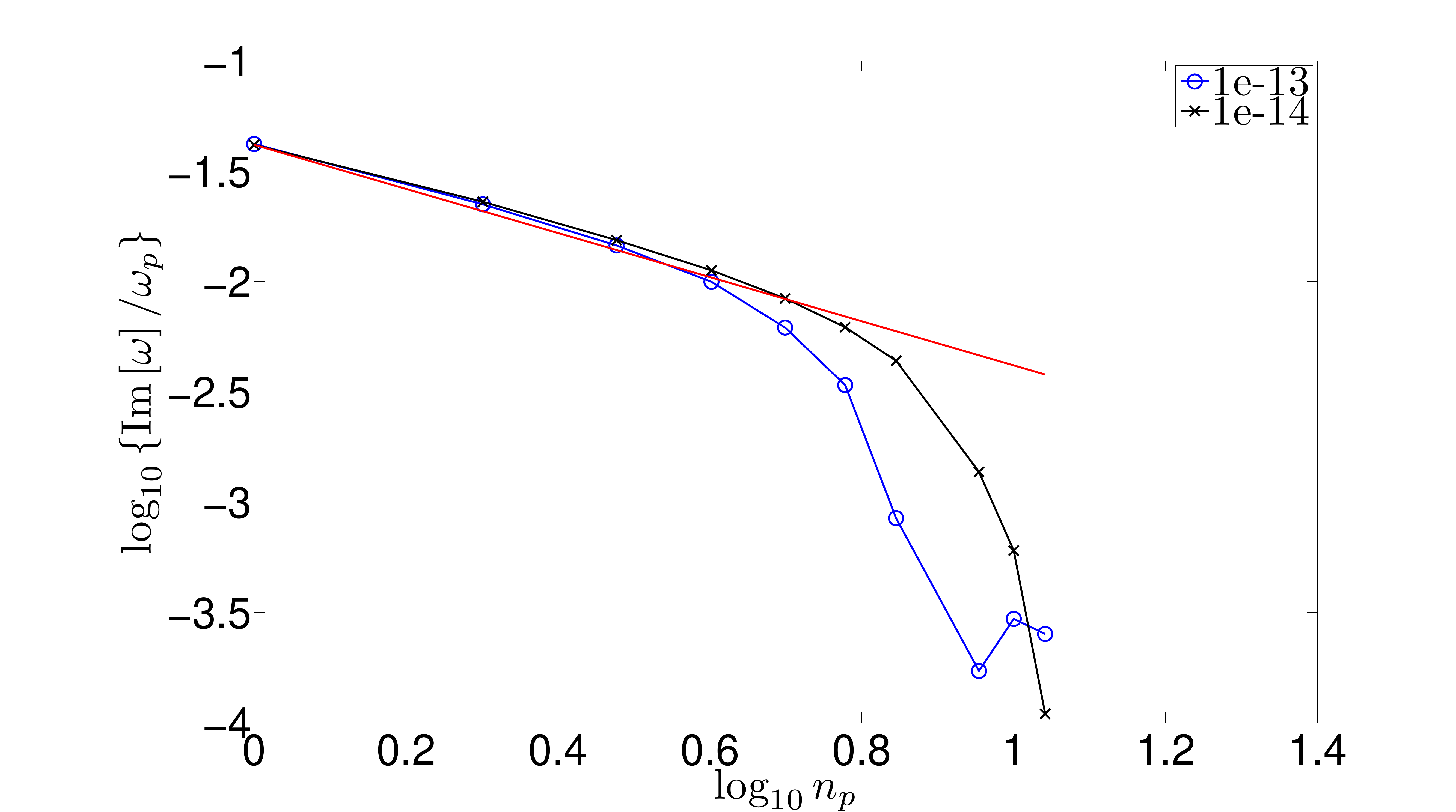} }
    \subfigure{\includegraphics[width=0.48\textwidth]{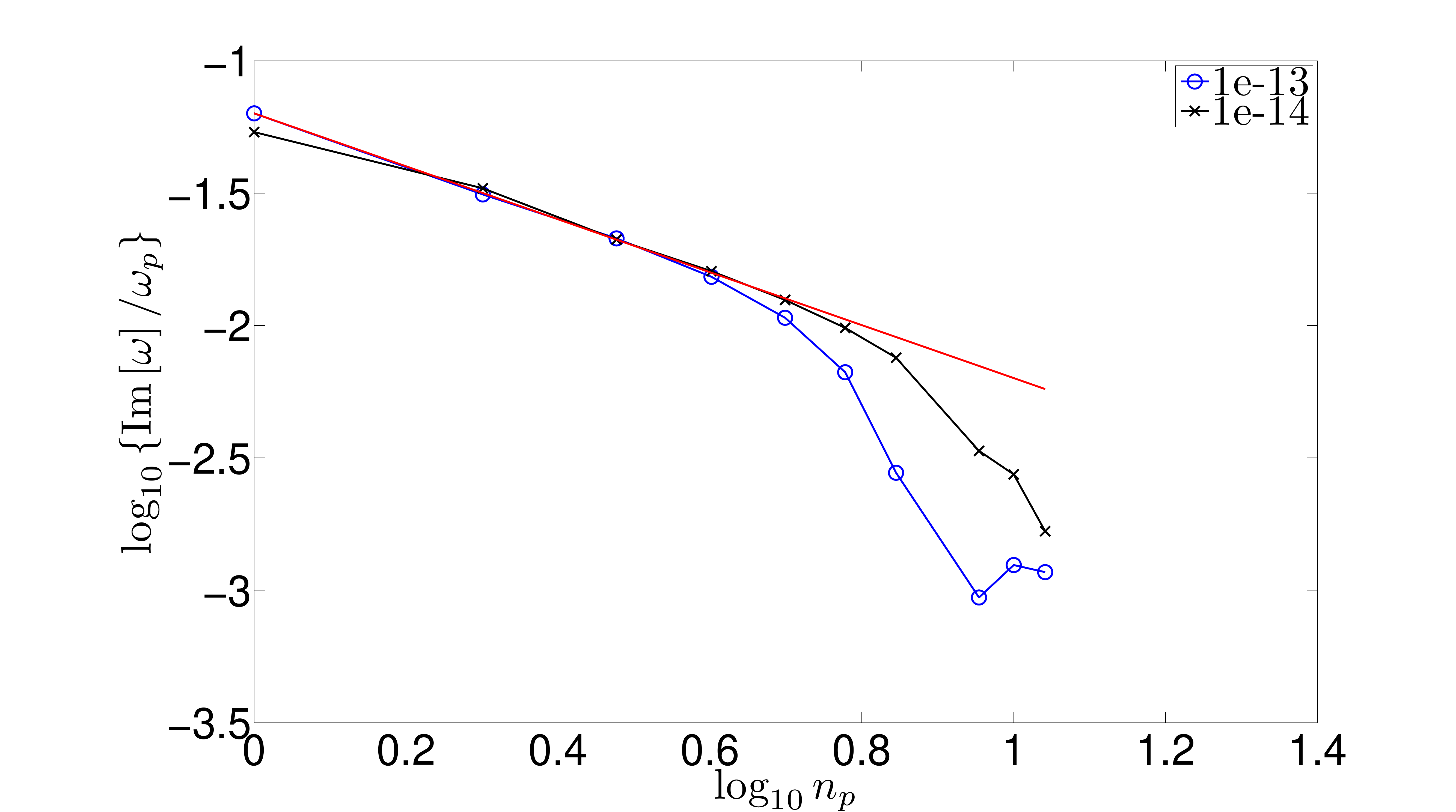} }
  \end{center}
  \caption{$\log_{10}$Im$\left[\omega/\omega_{p}\right]$
    vs. $\log_{10} n_{p}$ for the most unstable mode when $A=0.5$ and $0.8$
    respectively, for $\nu=10^{-13}$ and $10^{-14}$. The
    red line has a slope $-1$. The tail-off at larger $n_{p}$ is due
    to diffusion.}
  \label{galerkinpar5}
\end{figure}

In Fig.~\ref{galerkinpar6}, the growth rate for the
most unstable mode is shown to scale approximately
linearly with $A$ for $A\leq 1$. A slope of $1$ in this figure is predicted for
$A \ll 1$ if the instability is due to a parametric resonance, and
this appears to approximately hold for all $A$ in this range.

Our most important
result of this study is illustrated in Fig.~\ref{galerkinpar5}. This shows that the growth rate scales
inversely with the number of wavelengths within the domain (note that
this is after normalising by $\omega_{p}$). In this figure we
plot the logarithm of the growth rate versus $\log_{10} n_{p}$ for
$A=0.5$ and $0.8$, respectively. The slope is
always approximately equal to $-1$ for low $n_{p}$. The tail-off at larger $n_{p}$ is
due to diffusion, and arises because modes with smaller spatial
scales parametrically excite modes with even smaller spatial scales
(as a result of the theorem proved by \citealt{Hasselman1967}),
which are then more easily damped by diffusion. As we would
expect from this interpretation, the value of
$n_{p}$ at which diffusion dominates moves to smaller $n_{p}$ as $\nu$
is increased. The inverse dependence on $n_{p}$ that is present when
diffusion is unimportant is a key result. This suggests that although
parametric instabilities exist for any amplitude in the absence of
diffusion, in a
sufficiently large domain they become
unimportant. We discuss the relevance of this result to tidal
dissipation in \S~\ref{galerkindiscussion}.

\subsection{Eigenfunctions}

\begin{figure}
  \begin{center}
    \subfigure{\includegraphics[width=0.5\textwidth]{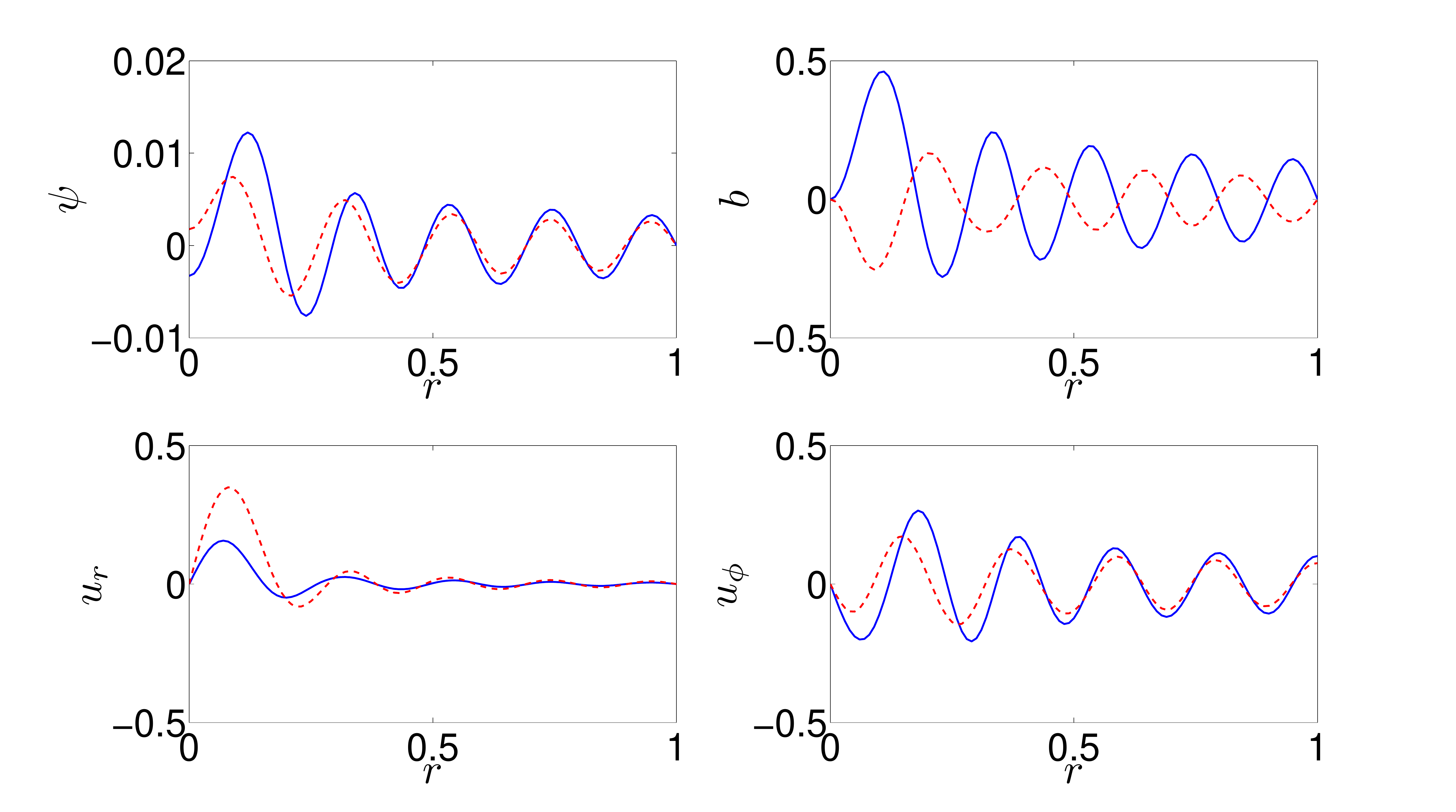} } \\
    \subfigure{\includegraphics[width=0.5\textwidth]{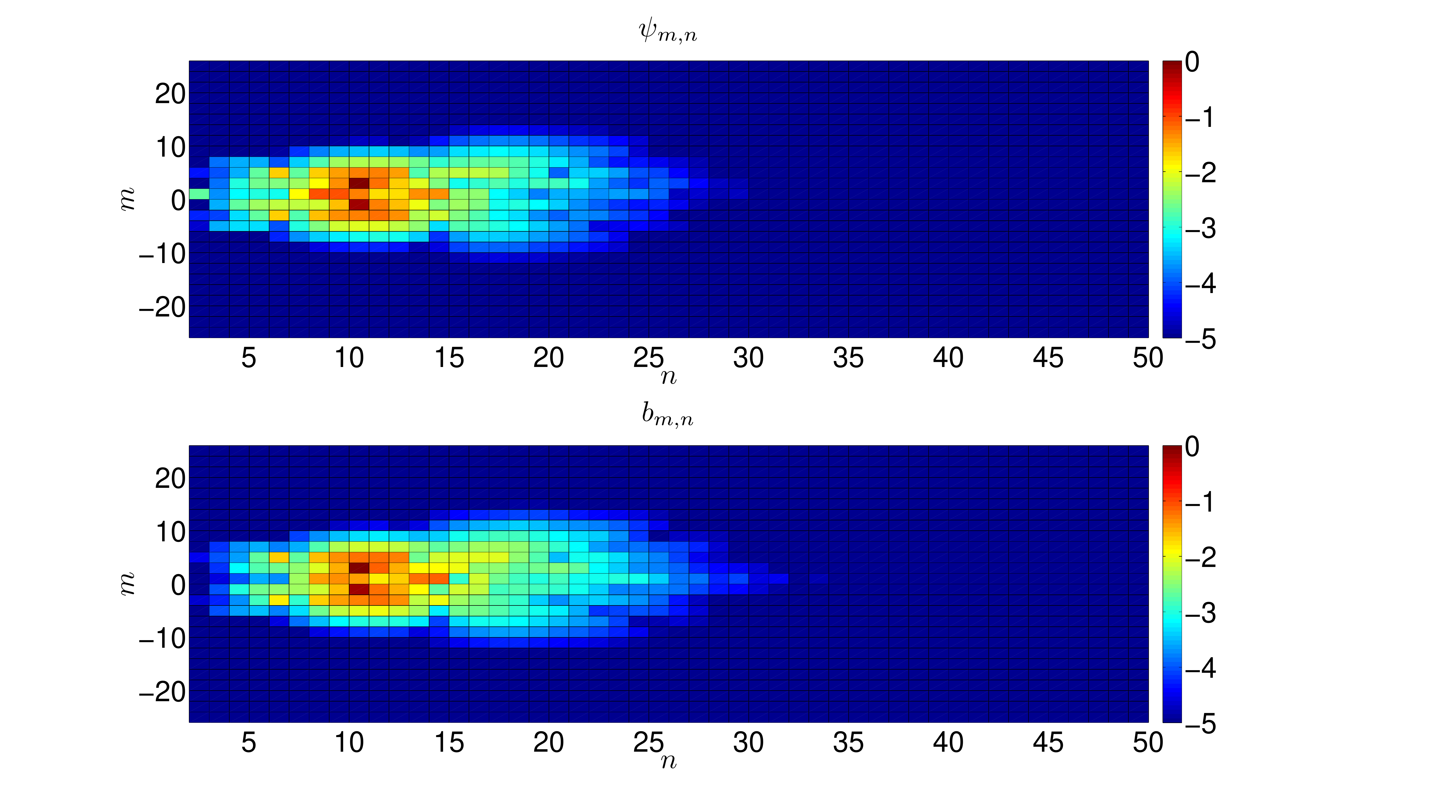} }
  \end{center}
  \caption{Top: Real (blue solid lines) and imaginary (red dashed
    lines) parts along $\phi=0$ of the spatial eigenfunctions for the most unstable mode for
    $A=0.1$, $n_{p}=2$, $\nu=10^{-13}$. The eigenfrequency is
    $\omega/\omega_{p} = 0.378 + 0.001i$. Bottom: Spectral space
    eigenfunction of the same mode. The colour scale represents
    $\log_{10} |\psi_{m,n}/\mathrm{max}\{\psi_{m,n}\}|$, and similarly
    for $b_{m,n}$.}
  \label{galerkinpar3}
\end{figure}

\begin{figure}
  \begin{center}
    \subfigure{\includegraphics[width=0.5\textwidth]{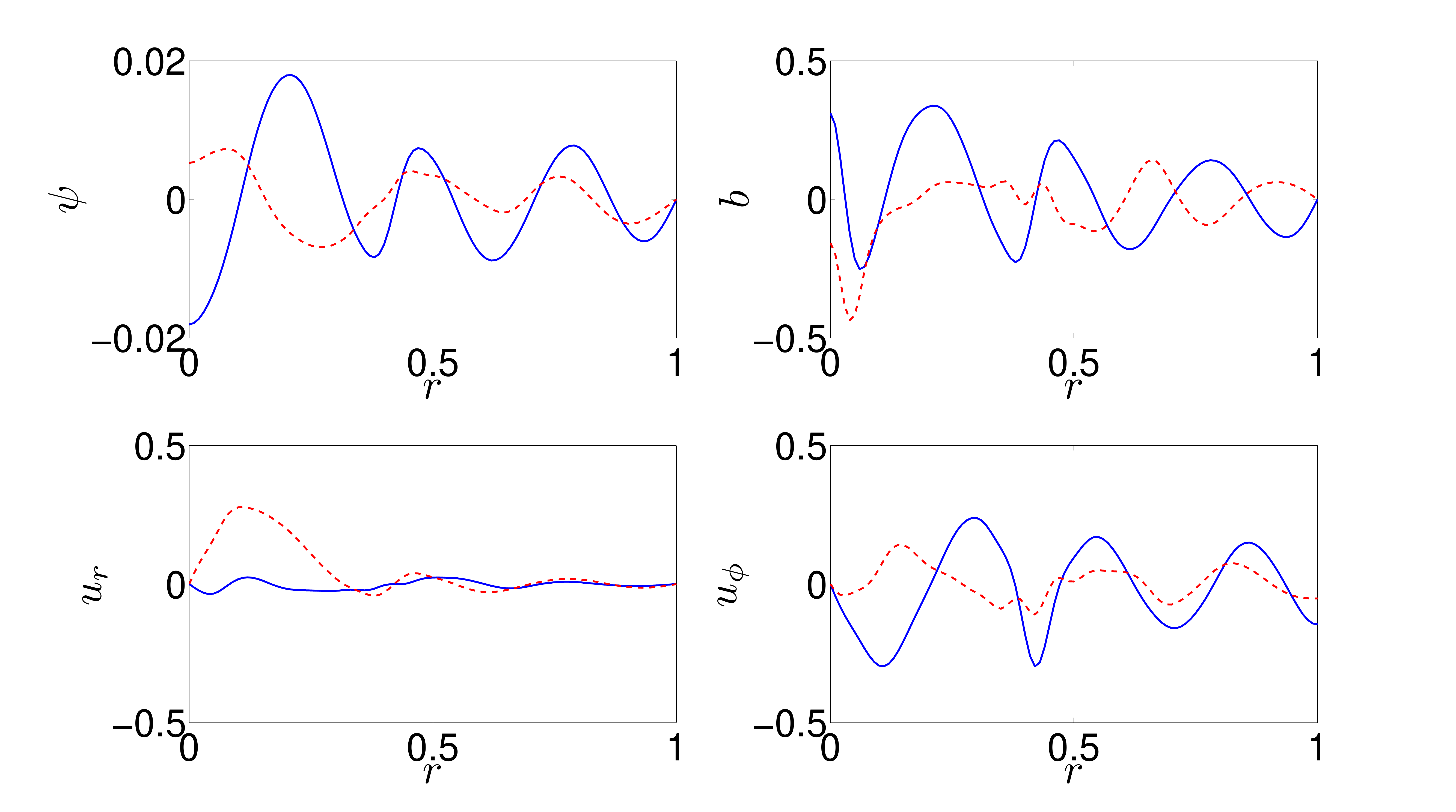} } \\
    \subfigure{\includegraphics[width=0.5\textwidth]{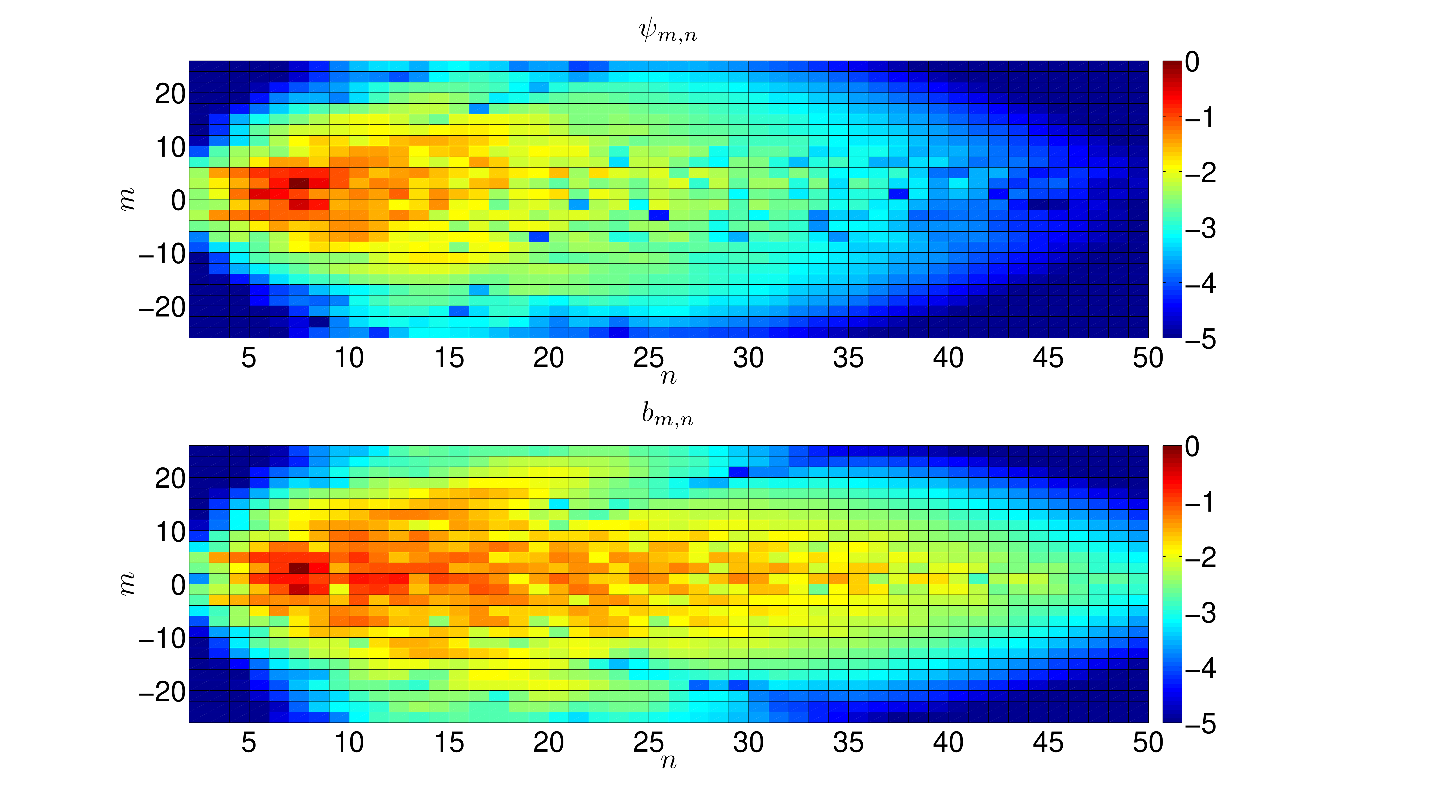} }
  \end{center}
  \caption{Top: Real (blue solid lines) and imaginary (red dashed
    lines) parts along $\phi=0$ of the spatial eigenfunctions for the most unstable mode with
    $A=1$, $n_{p}=2$, $\nu=10^{-13}$. The eigenfrequency is
    $\omega/\omega_{p} = 0.540 + 0.017i$. Bottom: Spectral space
    eigenfunction of the same mode. The colour scale represents
    $\log_{10} |\psi_{m,n}/\mathrm{max}\{\psi_{m,n}\}|$, and similarly
    for $b_{m,n}$.}
  \label{galerkinpar4}
\end{figure}

In the top panels of Figs.~\ref{galerkinpar3} and
\ref{galerkinpar4} we plot the real (blue solid lines) and imaginary
parts (red dashed lines) along $\phi=0$ of the
spatial eigenfunctions of the most unstable mode for
$A=0.1$ and $0.5$, respectively. We have taken $n_{p}=2$ and $\nu=10^{-13}$. 
These modes exist throughout the box and are confined by the boundary. They become more distorted from the
free wave modes as $A$ is increased towards unity (and in fact also for
$A>1$, apart from the localised modes), especially in the innermost wavelengths
of the primary wave. Note that the amplitude of the eigenfunctions is
 arbitrary since we are solving an eigenvalue problem.
The bottom panels of Figs.~\ref{galerkinpar3} and
\ref{galerkinpar4} show the spectral space eigenfunctions of the same unstable
modes. We have normalised $\psi_{m,n}$ and $b_{m,n}$ to their
maximum values and taken the base 10 logarithm of each component to
produce the figures. These show that growing modes for the chosen
values of $A$ are not simply a pair of free
wave modes which are excited by the primary wave. They contain
many $n$ and $m$ values localised around a particular $n$ and
$m$, and are therefore interacting strongly with the primary wave. Multiple
$n$ and $m$ values are involved even when $A=0.1$. For the value of $\nu$
adopted, these modes are well resolved, as is shown from the amplitude
decay of the spectral space eigenfunction, which occurs before the resolution limit
is reached in $n$ and $m$.

\subsection{Energetics of the instabilities}

When $A<1$, the isentropes are never overturned by the primary wave,
so a pure radial convective instability is not possible. However,
these instabilities could be driven by the free energy resulting from
the primary wave shear or entropy gradients, 
or a combination of the
two. In this section, we compute the spectral-space energy
contributions outlined in \S~\ref{spectralenergy}, for a sample of
growing modes in this amplitude range. We have confirmed that the growth
rate is accurately computed from Eq.~\ref{growthratepredictions}, to
within a few percent at most. In Table \ref{table1} we outline the
contributions to the growth rate from each term in
Eqs.~\ref{Energyeqs1}--\ref{Energyeqs3} for the most unstable mode for
$A=0.1,0.5$ and $1$ with $n_{p}=2$ and $\nu=10^{-13}$. The eigenfunctions for two of these are plotted
in Figs.~\ref{galerkinpar3} and \ref{galerkinpar4}. These examples are
illustrative of every unstable mode that exists when $A<1$ (and also
the non-localised modes that exist when $A>1$).

\begin{table}
\begin{center}
\begin{tabular}{l || c | c | r} 
 & $A=0.1$ & $A=0.5$ & $A=1$ \\ 
\hline
$\mathrm{Re}\left[\omega\right]$ & $0.065$ & $0.080$ & $0.093$ \\
$\mathrm{Im}\left[\omega\right]$ & $1.86 \times 10^{-4}$ & $1.97\times 10^{-3}$ &$3.00\times 10^{-3}$ \\
$ K$ & $3.28 \times 10^{-2}$ & $3.82 \times 10^{-2}$ & $3.70 \times 10^{-2}$\\ 
$ P$ & $3.25 \times 10^{-2}$ & $3.87 \times 10^{-2}$ & $4.16 \times 10^{-2}$\\
$ \mathcal{N}_{sw}$ & $-3.20 \times 10^{-7}$ & $-3.98\times 10^{-5}$& $-9.55 \times 10^{-5}$\\ 
$ \mathcal{N}_{bw}$ & $3.57 \times 10^{-5}$ & $3.76 \times 10^{-4}$ &$6.62 \times 10^{-4}$\\ 
$ F_{b}$ & $1.81 \times 10^{-5}$ & $2.04 \times 10^{-4}$& $3.51 \times 10^{-4}$\\
$ F_{\nu}$ & $-5.58 \times 10^{-6}$ & $-1.35 \times 10^{-5}$ & $-3.29 \times 10^{-5}$\\
$ F_{\kappa}$ & $-5.55 \times 10^{-6}$ & $-2.01 \times 10^{-5}$ &$-6.03 \times 10^{-5}$\\
\end{tabular}
\end{center}
\caption{Energy components of the most unstable mode for $A=0.1,0.5$ and
  $1$, with $n_{p}=2$ and  $\nu=10^{-13}$. Note that
  $\omega_{p}\approx0.17$.}
\label{table1}
\end{table}

Firstly, we note that the integrated kinetic and potential energy of
the modes are in approximate equipartition. A single wave can be
proved to be in exact equipartition (as is shown in Appendix \ref{equipartition}), so we would expect $
K \approx  P$ if these modes are parametrically
excited gravity waves with a single $n$ and $m$. That they are in
approximate equipartition and include many $n$ and $m$ components
indicates that these modes are the larger $A$ generalisations of the
parametrically excited free wave modes. The source of free energy driving these modes
is entirely the potential energy resulting from primary wave
entropy gradients. Somewhat surprisingly, the primary
wave shear contribution is much smaller, and actually stabilises the
modes. This instability converts primary wave potential energy
to disturbance potential energy, and then converts approximately half of this input
energy to the disturbance kinetic energy through the buoyancy flux
term. This process results in approximate equipartition between
$ K$ and $ P$.
Note that the entropy gradients in the primary wave are insufficient
to cause convective instability. These modes are driven by weaker
entropy gradients in radius and azimuth.

\section{Results for waves with $A>1$: the initial stages of wave breaking}
\label{breakingresults}

When $A>1$, the primary wave overturns the stratification during part
of its cycle. Our simulations in BO10 have shown that an instability
breaks the wave within a few wave periods once this first
occurs. The initial stages of this breaking process are examined in
this section by choosing $A>1$. To resolve convectively unstable modes
with the adopted values of $L_{n}$ and $L_{m}$, we require the size of the overturning
region to be sufficiently large. Since overturning occurs only at the point $r=0$ when $A=1$, 
this necessitates choosing values of $A$ larger than unity to capture
such instabilities. We are interested in instabilities which act to
break the waves in an (effectively) unbounded domain (the central
regions of the RZ of a solar-type star), therefore the appropriate
unstable mode should not rely on the boundaries for confinement, and
should be localised within the innermost wavelength of the primary
wave. This is because the presence of confining boundaries is
artificial, and is imposed to specify the problem. 
With this in mind, we now discuss the results of our stability
analysis for waves with $A>1$.

\begin{figure}
  \begin{center}
    \subfigure{\includegraphics[width=0.48\textwidth]{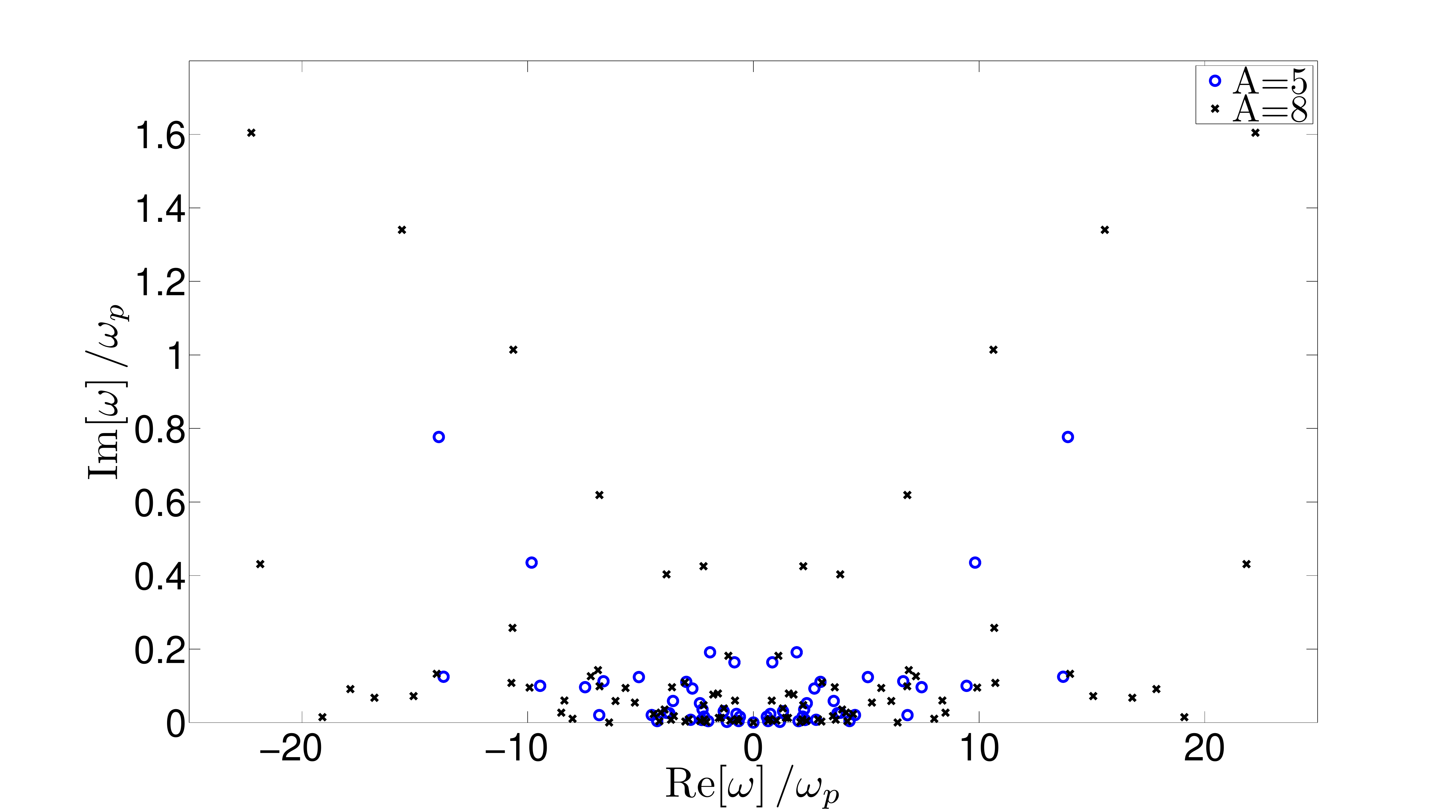} }
\end{center}
  \caption{Distribution of unstable eigenvalues on the complex
    frequency plane for $n_{p}=2$ and various $A$ with
    $\nu=10^{-13}$.
  } 
  \label{galerkinbreak1}
\end{figure}

\begin{figure}
  \begin{center}
    \subfigure{\includegraphics[width=0.48\textwidth]{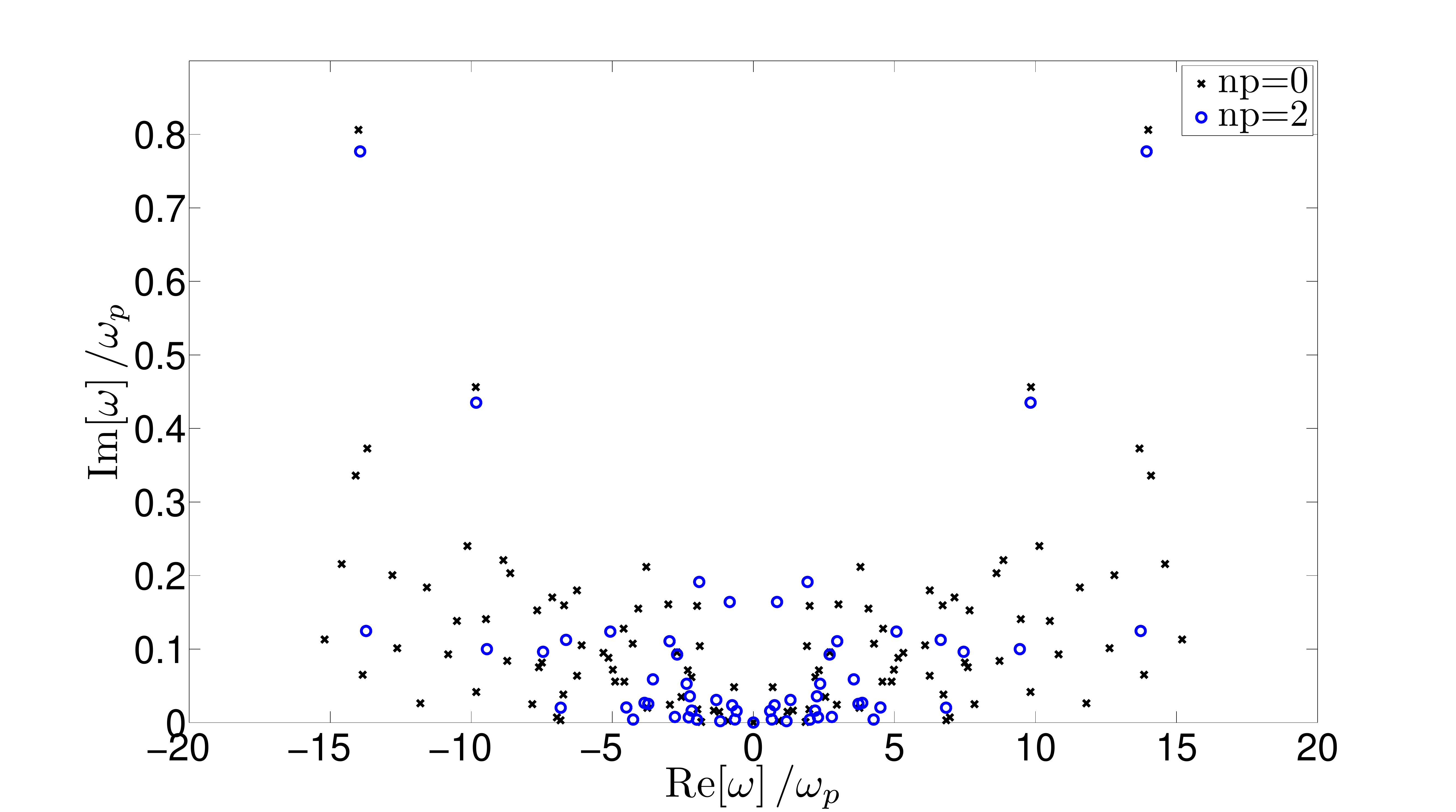} }
\end{center}
  \caption{Distribution of unstable eigenvalues on the complex
    frequency plane for $A=10$ and various $n_{p}$, with $\nu=10^{-13}$.} 
  \label{galerkinbreak2}
\end{figure}

The eigenvalues of the unstable modes displayed on the
complex plane are shown in Fig.~\ref{galerkinbreak1} for $A=5$ and
$A=8$, both with $n_{p}=2$ and $\nu=10^{-13}$. The most unstable modes
are located on distinct branches, which stand above the continuation
of the modes that exist when $A<1$. From studying their
eigenfunctions, we find that the modes on the branches are localised disturbances,
unlike those below the main branches. The modes on the branches could
therefore represent the type of mode that breaks the primary
wave (this is discussed further in the next subsection). 
These branches extend further from the
origin the greater the value of $A$. In Fig.~\ref{galerkinbreak2} we plot the unstable modes
for $A=5$ for two values of $n_{p}$. This shows that the
unstable modes on the branches do not move around significantly as $n_{p}$ is
varied. They therefore depend only weakly on the location of the outer boundary. Note,
however, that the growth rate becomes smaller as we go to larger $n_{p}$
because of the increasing importance of diffusion.

Note that the largest frequency of some growing modes is larger than
the maximum buoyancy frequency 
$N_{max}=1$ (which corresponds with $1/\omega_{p}
\sim 6$ in the units of this figure, for $n_{p}=2$). The nonzero
frequencies of the modes in this frame indicate that they are oscillatory, and are non-steady.
In addition, the growth rates of the most unstable modes are sufficiently fast compared
with the primary wave frequency that the
instability grows within several wave periods after onset. This is
consistent with the results of our simulations described in BO10.

\begin{figure}
  \begin{center}
    \subfigure{\includegraphics[width=0.48\textwidth]{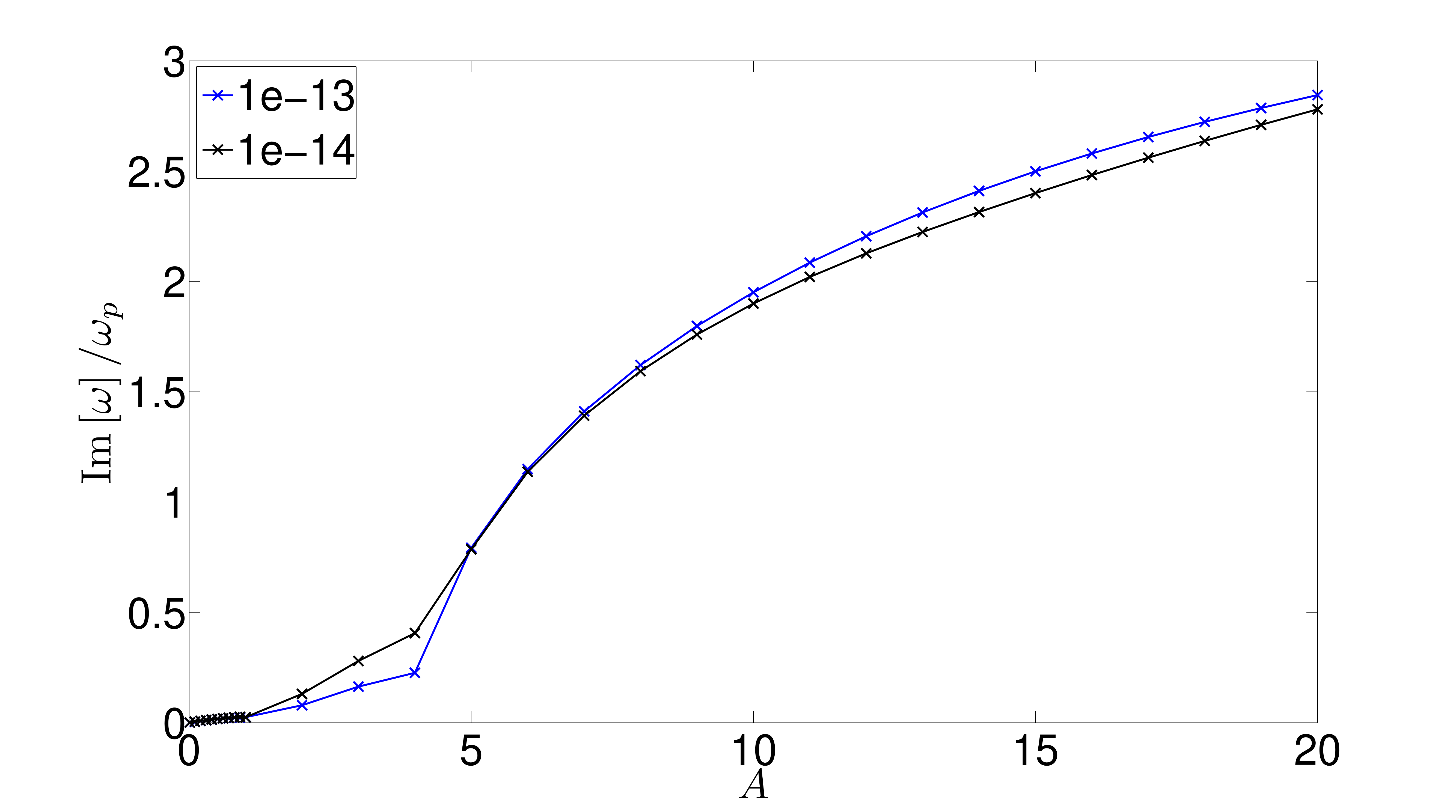} } 
  \end{center}
  \caption{Im$\left[\omega\right]$
    vs.~$A$ for the most unstable mode for $n_{p}=2$, $\nu=10^{-13}$
    and $10^{-14}$. The instability grows within a
    primary wave period when $A\gtrsim 5$, which is when the localised
    modes begin to appear.}
  \label{galerkinbreak3}
\end{figure}

The growth rate of the most unstable modes for a given
$n_{p}$ increases with $A$ as illustrated in Fig.~\ref{galerkinbreak3} for
$n_{p}=2$, where curves for $\nu = 10^{-13}$ and $10^{-14}$ have
been plotted. There is an approximate square root
dependence for $A\gtrsim 4$. If the instability is driven by convectively unstable entropy
 gradients, then we might expect
 \begin{eqnarray}
\nonumber
  \mathrm{Im}\left[\omega^{\prime}\right] \hspace{-0.2cm} &\lesssim&
  \hspace{-0.2cm} \sqrt{\mathrm{max}\left[-N^{2}\right]} \\ \nonumber
\hspace{-0.2cm} &=& \hspace{-0.2cm}\left(\mathrm{max}\bigg\{-r^{2} - \frac{2
       A}{k_{2,n_{p}}}r\left(J_{1}(k_{2,n_{p}}r) \right. \right. \\
   \nonumber && \left. \left. \hspace{2.5cm}
-J_{3}(k_{2,n_{p}}r)\right)\cos 2\phi\bigg\}\right)^{1/2}.
 \end{eqnarray}
Thus, for large $A$ the growth rate should scale with the square root of the
primary wave amplitude. This behaviour is not observed
when $1 \lesssim A \lesssim 4$. In this range, the square root
dependence may not be exhibited 
partly because there is
insufficient resolution to accurately capture the modes
that contribute to breaking since the overturning region is small
compared with the box size. We have noticed that the growth
rate does not significantly depend on $\nu$ (and therefore the
resolution), for the most unstable modes, except in the
range $1 \lesssim A \lesssim 4$, which supports this explanation.

\begin{figure}
  \begin{center}
    \subfigure{\includegraphics[width=0.48\textwidth]{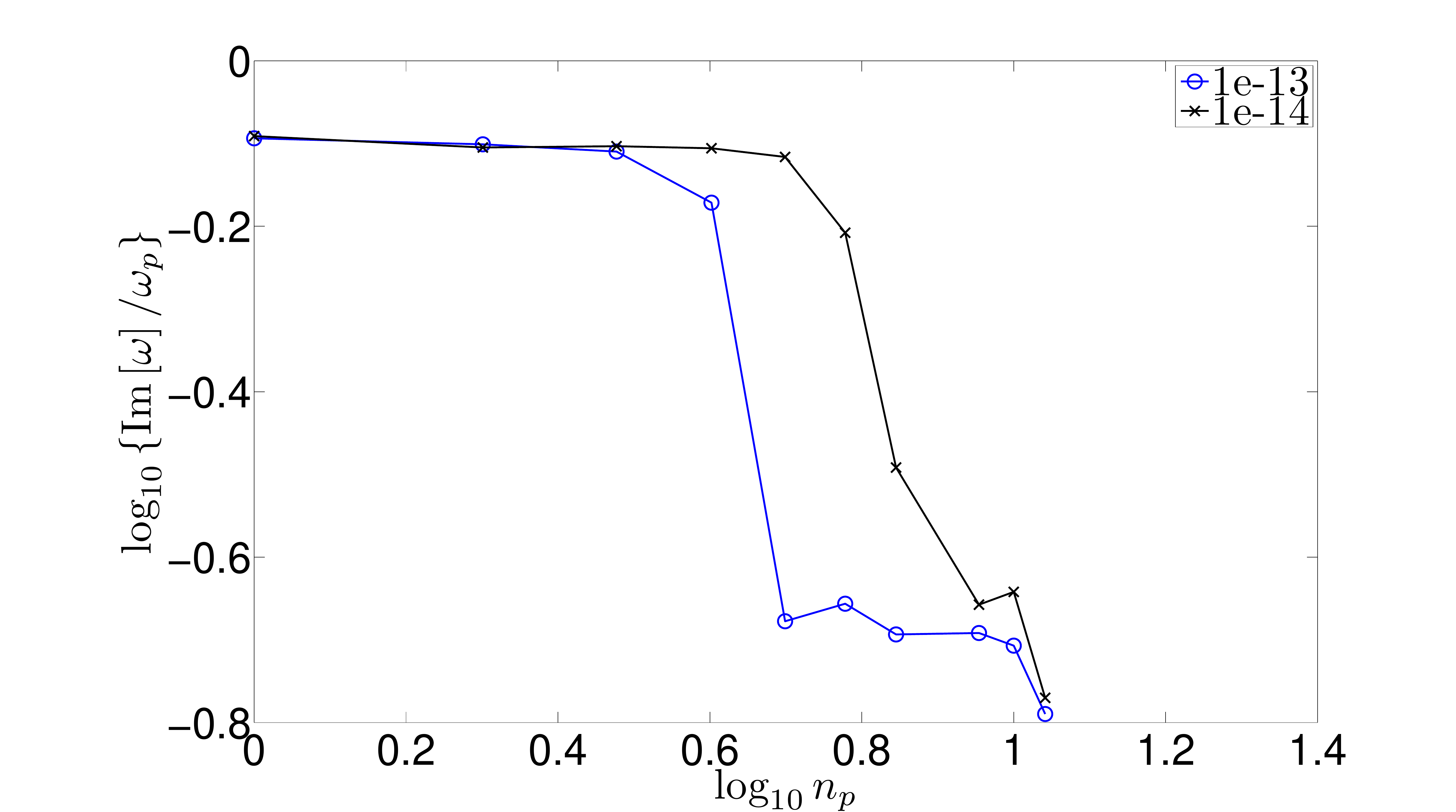} }
  \end{center}
  \caption{Im$\left[\omega\right]$
    vs.~$n_{p}$ for the most unstable localised modes when $A=5$ and 
    $\nu=10^{-13}$ and $10^{-14}$.}
  \label{galerkinbreak7}
\end{figure}

The behaviour of the growth rate on $n_{p}$ is
illustrated in Fig.~\ref{galerkinbreak7} for the most unstable mode when
$A=5$, for two values of $\nu$. For
large $n_{p}$, the unstable modes have sufficiently small spatial
scales for diffusion to become important, so we expect a tail-off at
large $n_{p}$. The important point that can be taken from
this figure is that the (normalised) growth rate of the localised
modes on the branches does not
depend on the number of wavelengths within the domain, for
modes that are not strongly affected by diffusion. This
means that the instability can be important in a large domain, such as
a the RZ of a solar-type star, which contains many primary
wavelengths. Note that this is very different to the behaviour found
for the excited modes when $A<1$, as shown in Fig~\ref{galerkinpar5}.

\subsection{Eigenfunctions}

 \begin{figure}
   \begin{center}
     \subfigure{\includegraphics[width=0.5\textwidth]{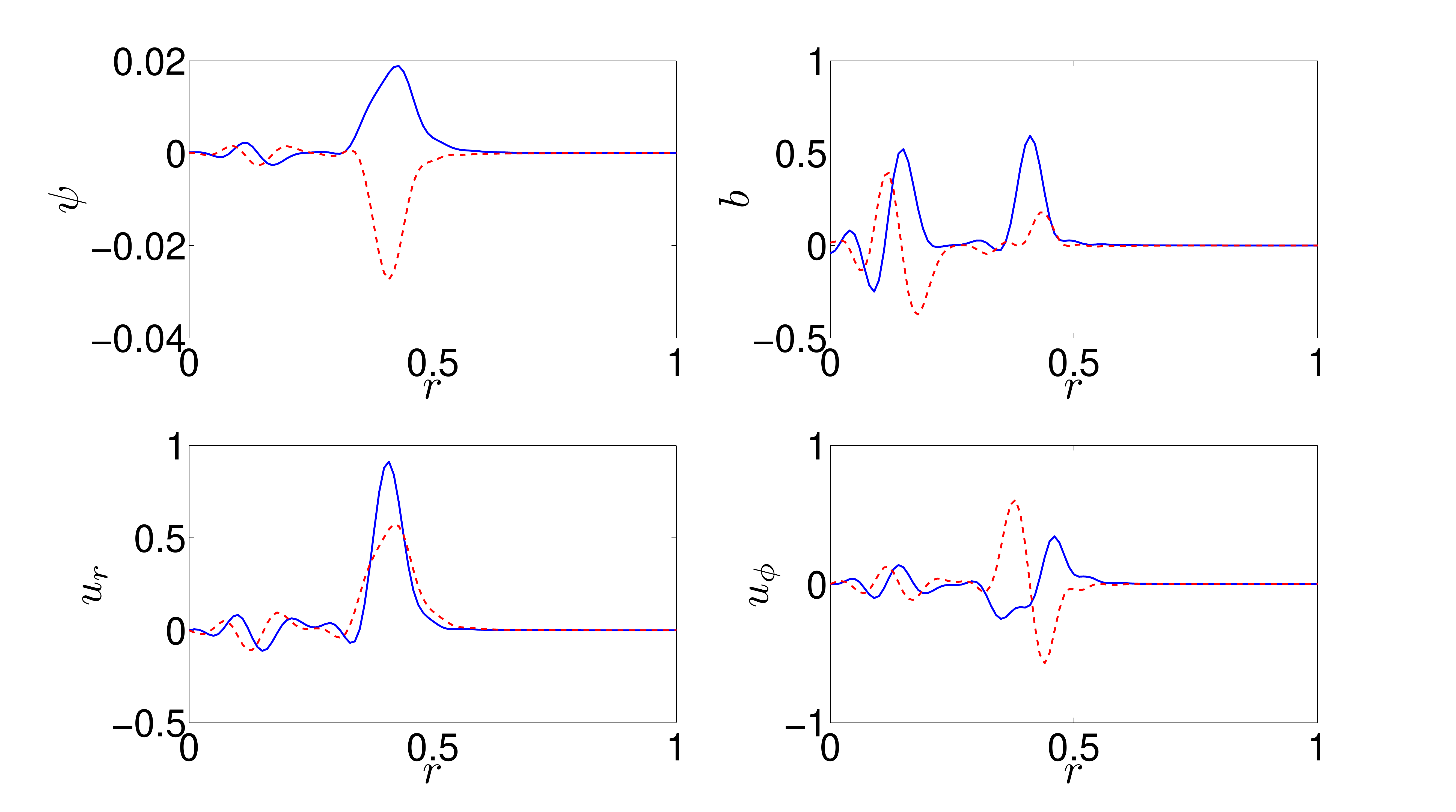} } \\
     \subfigure{\includegraphics[width=0.5\textwidth]{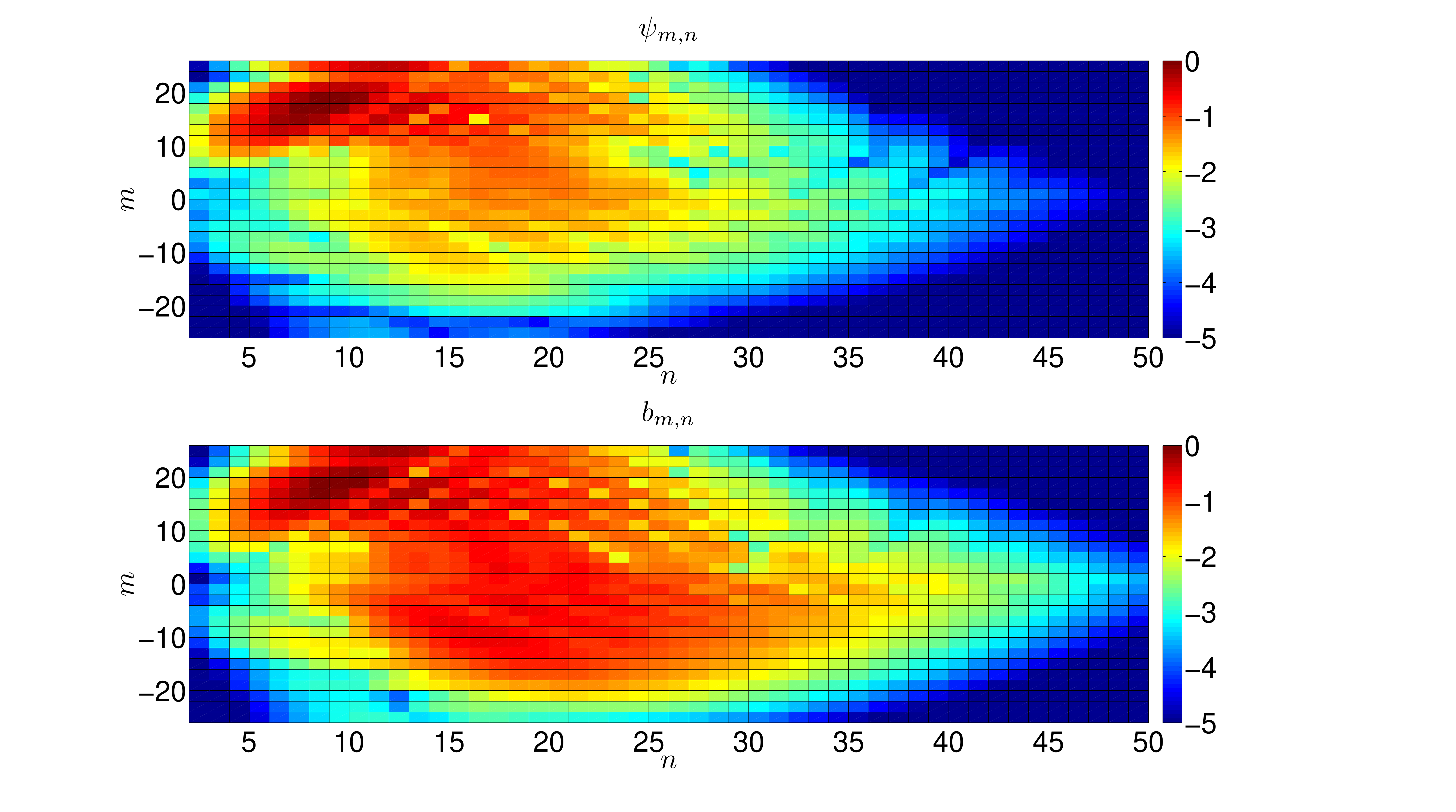} }
   \end{center}
   \caption{Top: Real (blue solid lines) and imaginary (red dashed
    lines) parts along $\phi=0$ of the spatial eigenfunctions for the most unstable mode with
    $A=5$, $n_{p}=2$, $\nu=10^{-13}$. The eigenfrequency is
    $\omega/\omega_{p} = 9.68 + 0.302i$. Bottom: Spectral space
    eigenfunction of the same mode. The colour scale represents
    $\log_{10} |\psi_{m,n}/\mathrm{max}\{\psi_{m,n}\}|$, and similarly
    for $b_{m,n}$.}
   \label{galerkinbreak4}
 \end{figure}

 In the top panel of Fig.~\ref{galerkinbreak4} we plot the
 real (blue solid lines) and imaginary parts (red dashed lines) along
 $\phi=0$ of the
 spatial eigenfunctions of the most unstable mode for $A=5$,
 $n_{p}=2$ and $\nu=10^{-13}$. The spectral-space eigenfunction of this mode is plotted in
 the bottom panel. The mode is strongly nonlinearly interacting with the primary
 wave, as shown by the number of different $n$ and $m$ values that
 appreciably contribute. The contribution to the eigenfunction is nonzero, but not maximal,
 near $|m|=L_{m}-1$, and is negligible at $n=L_{n}$, which indicates that this
 mode is adequately resolved.
 The eigenfunction is spatially localised within the innermost wavelengths
 of the primary wave. Each of the several most unstable
 modes in the range $5\leq A\leq 10$ which lie on the branches in
 Fig.~\ref{galerkinbreak1} are localised modes, and have qualitatively different form to the
 type of modes that exist below the branches, which are a continuation
 of the modes that exist when $A<1$. As we go to larger $A$ for the same value of $\nu$, the most unstable mode
utilises an increasing number of $n$ and $m$ values up to the
resolution limit. This means that to adequately resolve the modes we would have
to either increase the resolution or the value of $\nu$.

\begin{figure}
  \begin{center}
    \subfigure{\includegraphics[width=0.5\textwidth]{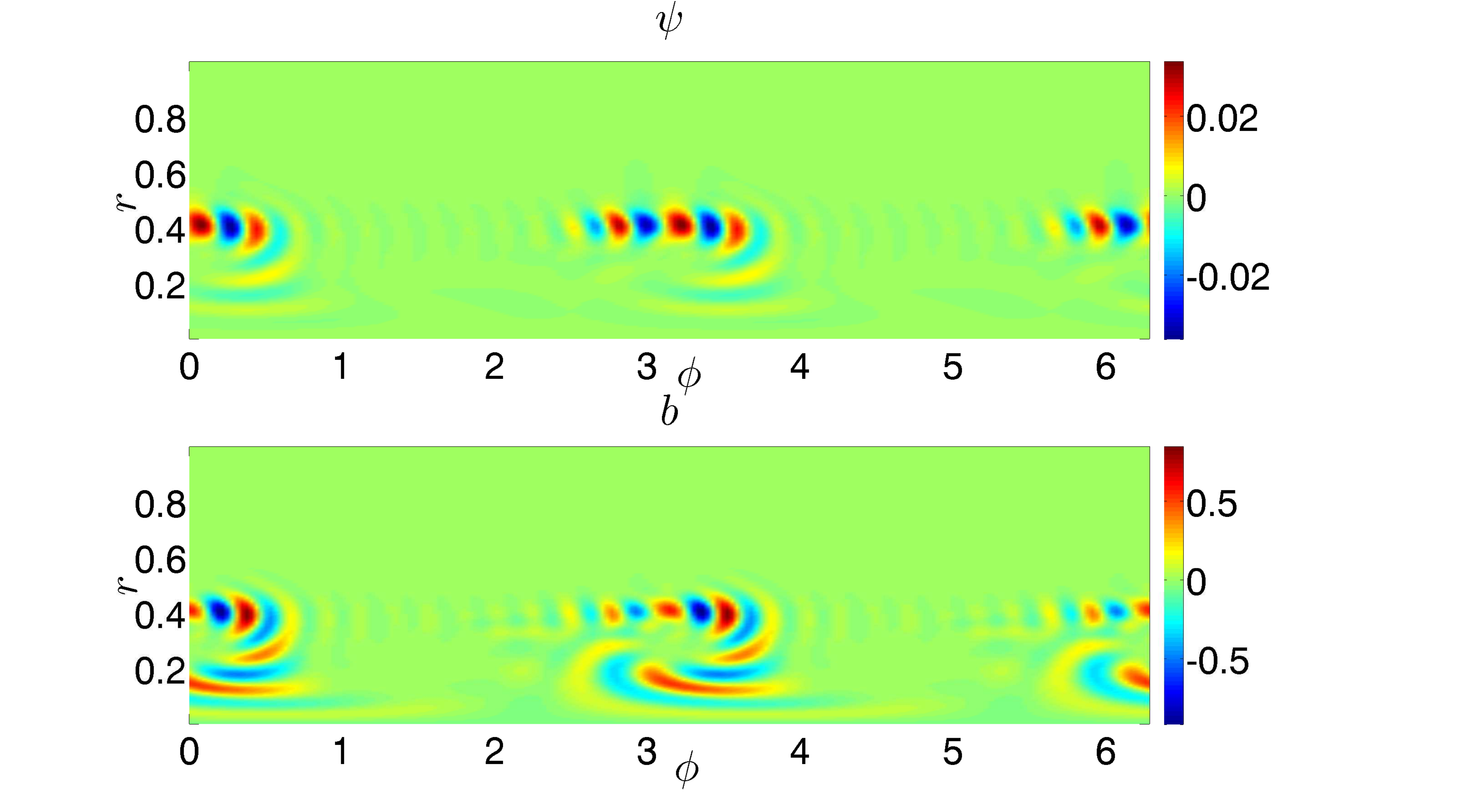} } \\
    \subfigure{\includegraphics[width=0.5\textwidth]{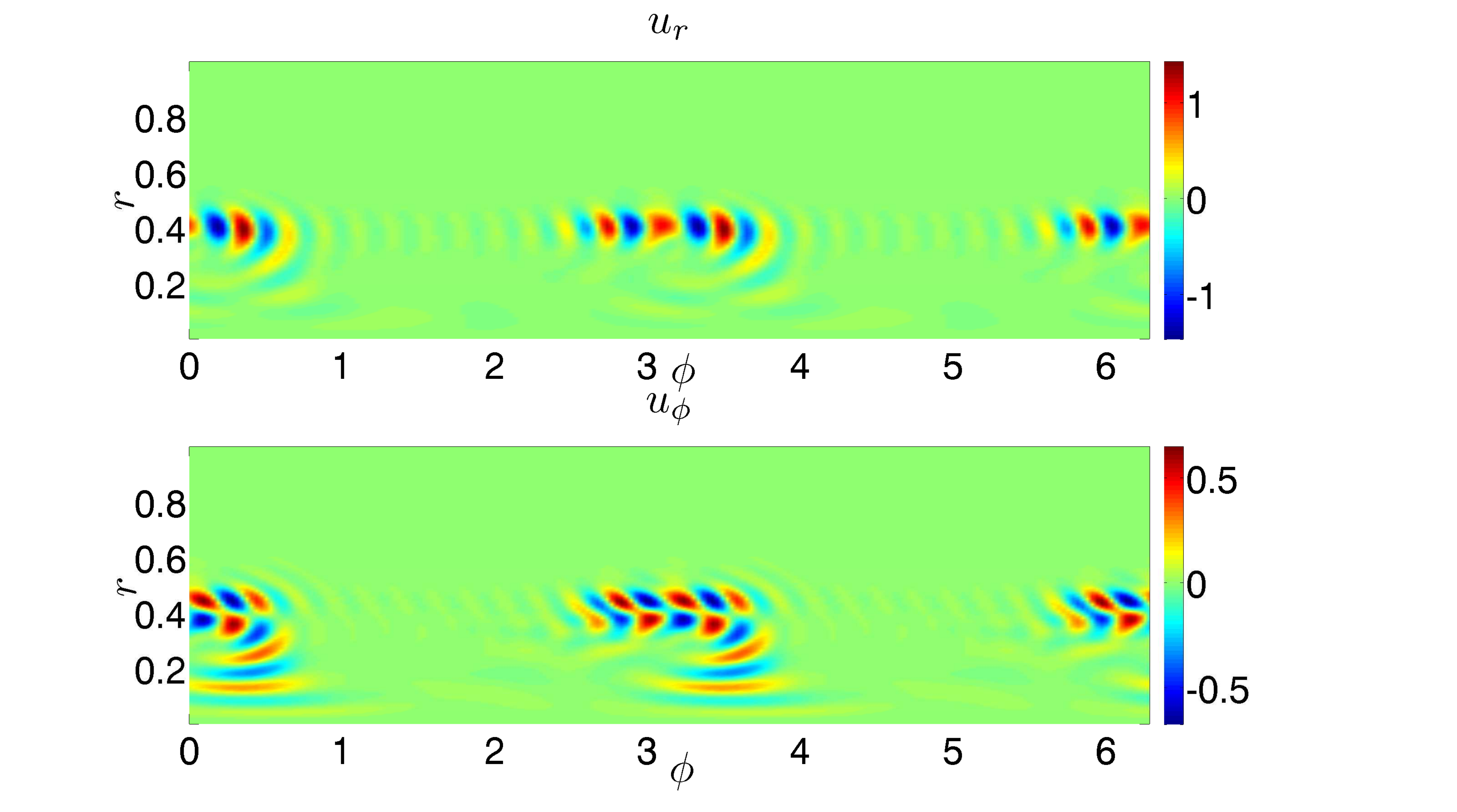} }
  \end{center}
  \caption{2D Spatial eigenfunction of the most unstable mode for $A=10$,
    $n_{p}=2$ and $\nu=10^{-13}$, plotted on the $(\phi,r)$-plane.
}
  \label{galerkinbreak5}
\end{figure}

\begin{figure}
\begin{center}
  \subfigure{\includegraphics[width=0.45\textwidth]{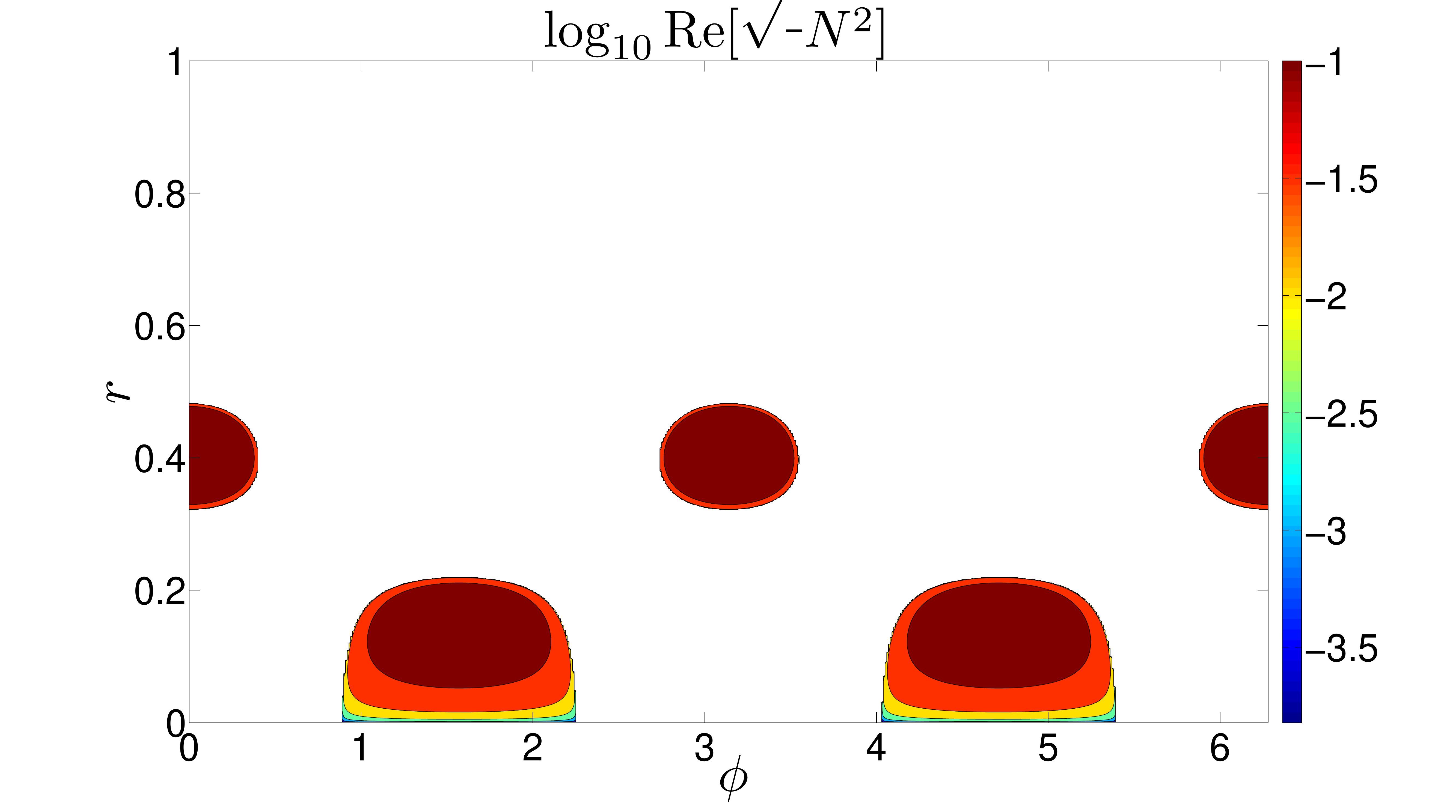} }
 \end{center}
  \caption{Unstable region for $A=10$, $n_{p}=2$. This can be compared with
    Fig.~\ref{galerkinbreak5}.}
  \label{galerkinbreak6}
\end{figure}

The components of the spatial eigenfunction of the most unstable mode when $A=5$,
$n_{p}=2$ and $\nu=10^{-13}$, is plotted in
Fig.~\ref{galerkinbreak5} on the $(\phi,r)$-plane, to
further illustrate the spatial dependence of this mode. In
Fig.~\ref{galerkinbreak6} we plot
the region of negative $N^{2}$ for the same primary wave. A comparison
of these figures makes clear that the eigenfunction is primarily localised within
the regions made convectively unstable by the primary wave entropy
perturbation. This adds further evidence to the conjecture that the
instability is convective. We also find that any unstable mode on the branches of
Fig.~\ref{galerkinbreak1} that are excited when $5\leq A\leq 10$ are similarly localised and have a qualitatively similar appearance
 to the eigenfunction plotted in Fig.~\ref{galerkinbreak5}.

\subsection{Energetics of the instabilities}

In this section, we compute the spectral space energy contributions
outlined in \S~\ref{spectralenergy} for a representative sample of the
localised growing modes that exist when $A>1$. We have confirmed that the growth
rate is accurately computed from Eq.~\ref{growthratepredictions}, to
within a few percent for the modes
considered in this analysis. However, it must be noted that the most
unstable mode when $A\gtrsim 5$ is typically not fully resolved with
our adopted resolution and $\nu$, in that there is nonzero power in the highest $n$ and $m$
values. This can lead to errors in the energy analysis typically of
order $10-30 \%$, so we leave these modes out of this analysis, and
only choose those that are adequately resolved\footnote{A
    small number of higher resolution calculations with $L_{m}$ values
    up to $45$
have been performed to fully resolve these
modes. These calculations have confirmed that although the numerical
values for several quantities may differ
slighty for the most poorly resolved modes, our main results are not dependent on resolution.} for Table
\ref{table2}. In this table, we outline the
contributions to the growth rate from each term in
Eqs.~\ref{Energyeqs1}--\ref{Energyeqs3} for several unstable modes
that exist when $A\gtrsim 5$, each with $n_{p}=2$ and
$\nu=10^{-13}$. The eigenfunction corresponding to 
the first of these is plotted in Figs.~\ref{galerkinbreak4} and \ref{galerkinbreak5}.

\begin{table}
\begin{center}
\begin{tabular}{l || c | r} 
 & $A=5$ & $A=7$ \\ 
\hline
$\mathrm{Re}\left[\omega\right]$ & $1.67$ & $2.40$  \\
$\mathrm{Im}\left[\omega\right]$ & $0.052$ & $0.134$\\
$ K$ & $6.95 \times 10^{-2}$ & $3.37 \times 10^{-2}$ \\ 
$ P$ & $1.73 \times 10^{-2}$ & $1.02 \times 10^{-2}$ \\
$ \mathcal{N}_{sw}$ & $-5.50\times 10^{-3}$ & $-1.57 \times 10^{-3}$ \\ 
$ \mathcal{N}_{bw}$ & $2.05 \times 10^{-2}$ & $1.55 \times 10^{-2}$  \\ 
$ F_{b}$ & $1.82 \times 10^{-2}$ & $1.22 \times 10^{-2}$\\
$ F_{\nu}$ & $-2.10 \times 10^{-3}$ & $-1.33 \times 10^{-3}$  \\
$ F_{\kappa}$ & $-0.517 \times 10^{-3}$ & $-0.80 \times 10^{-3}$ \\
\end{tabular}
\end{center}
\caption{Energy components of the most unstable mode for $A=5$ and $7$, with $n_{p}=2$ and $\nu=10^{-13}$. Note that $\omega_{p}\approx0.17$.}
\label{table2}
\end{table}

As in the case of the modes that exist when $A<1$, the instability is
driven by the free energy associated with primary wave entropy
gradients, as is shown by the fact that $
\mathcal{N}_{bw}$ is the dominant contribution to the growth. 
This is indeed what would be expected of a convectively driven
instability. In addition, the primary wave shear is much weaker and tends to stabilise
the modes. Unlike the modes that exist when $A<1$, we
do not necessarily have $ K \approx  P$,
and examples have been found that do and do not satisfy approximate
equipartition, so these modes do not all appear to be gravity wave-like,
unlike the parametrically excited modes that exist when $A<1$.

The growth rates are always $\lesssim
\mathrm{Re}\left[\sqrt{-N^{2}}\right]$, which is expected to be an
upper limit if the instability is convective. The negative
contribution of shear, as well as hyperdiffusion mean that the modes
that we have calculated have somewhat smaller growth rates that this
simple estimate would predict. The route of energy transfer that
drives the instability is the same as for the parametric
instabilities discussed in the previous section.

\section{Summary and discussion of results}
\label{galerkindiscussion}

In the previous two sections we have analysed the instabilities that
exist when $A<1$ and $A>1$. 

\subsection{Wave breaking}

When $A>1$ we have identified a class
of localised modes that are driven by convectively unstable entropy gradients in
the primary wave. These modes exist in the absence of an outer
boundary, and are very likely to have initialised the wave breaking
process in the simulations presented in BO10 and B11. Subsequent stages in the breaking process are not studied
using this stability analysis because these would involve nonlinear
interactions between the perturbations to the wave, which we neglected
in our weakly nonlinear approach. The instability growth time is of
the order of a primary wave period, which is in agreement with the
wave breaking times observed in our simulations.

\subsection{Parametric instabilities}

When $A<1$, there exist pairs of parametrically excited modes driven by
(convectively stable) primary wave entropy gradients, with wave shear
playing a subordinate stabilising role. These modes exist because of
their confinement by the outer boundary. Our most important result
regarding these modes is the inverse dependence of the growth rate on
$n_{p}$. This can be explained by considering the relative time the
primary wave spends in the innermost regions, where its nonlinearity
is strongest. The fraction of the total wave propagation time spent in
the innermost regions, where growing modes are excited by the
nonlinearity, scales with $k_{2,n_{p}}^{-1} \propto n_{p}^{-1}$, so the
growth rate should scale also with $n_{p}^{-1}$. Combining this with
our observation that the growth rate increases approximately linearly
with wave amplitude, we can write
\begin{eqnarray}
\label{parametricgrowthrate}
\mathrm{Im}\left[\omega \right] \propto \frac{A}{n_{p}},
\end{eqnarray}
for the modes excited when $A<1$.

Our simulations in BO10 and B11 did not
show any instabilities acting on the waves when $A<1$. This can be
neatly explained from Eq.~\ref{parametricgrowthrate} in the limit as
$n_{p}\rightarrow \infty$, where the growth rate tends to zero. This
limit is appropriate since the waves have effectively no outer
boundary in the simulations, because we damp the waves before they
reach the boundaries of the computational domain. The fact that we
observed no instabilities in the simulations is therefore consistent with this
stability analysis, and is not a consequence of
limited run time or insufficient spatial resolution.

\subsection{Comparison with the plane IGW problem}
\label{comparisonIGW}
This problem has some important differences with the case of a plane IGW
in a uniform stratification. For that problem, as we discussed in the
introduction to this paper, parametric instabilities act for any
$A$, and always result in instability in the absence of diffusion. In
addition, \cite{LombardRiley1996} find that the presence or absence of
isentropic overturning does not seem to play a dominant role in, and
is not the cause of, the instability\footnote{However, it is possible
  that in their calculations they have insufficient resolution to be
  able to resolve any localised convectively unstable modes. If they
  go to larger $A$ than the maximum they consider of $1.1$, and/or consider
  larger resolutions, such localised modes may start to appear.}. This
is different from our problem, where we find that overturning
results in the presence of a different class of localised modes, that
are excited in the convectively unstable regions of the wave. However,
we do find that the source of free energy driving the instability,
which is the free energy
  associated with primary wave entropy gradients, is
the same whether $A<1$ or $A>1$. It is
true that parametric instabilities exist for any $A$ in our problem,
like in the plane IGW problem, but these become unimportant in a large domain because the
nonlinearity is spatially localised in the innermost wavelengths. This
is different from the plane IGW problem, in which the nonlinearity is
important everywhere in the wave.

The importance of overturning in our case could be because the primary
wave shear does not drive the modes, and in fact typically acts to
stabilise them. In the plane IGW problem, instabilities for any $A$
are driven by a combination of $\mathcal{N}_{sw}$ and
$\mathcal{N}_{bw}$, whereas in our problem
instabilities are always driven solely by $\mathcal{N}_{bw}$. \cite{Koudella2006}
  have also found $\mathcal{N}_{bw}$ to be the dominant energy source driving
  the instability of a (convectively stable) plane propagating IGW in 2D. They adopted a resonant triad
  model, which is valid for small wave amplitudes, to predict such a
  result. However, these calculations and ours were in two dimensions, so it remains to be seen
  whether shear would remain unimportant for our problem in 3D.

From the results of their stability analyses, \cite{LombardRiley1996}
and \cite{SonmorKlaassen1997} state that wave stability is a
three-dimensional problem. This might suggest that the picture we have
outlined could differ in 3D. However, the simulations performed in
B11 show a strong similarity with the 2D results in BO10.
Performing a similar stability analysis in 3D would be somewhat
involved, and would be restricted to studying the stability of small
amplitude waves, because the wave solution is not exact in 3D. It
would be possible to calculate higher order terms to the solution,
which would make it valid for larger $A$, and then perform a stability
analysis of this wave. Without performing such an analysis, it is
difficult to quantify the importance of three-dimensional effects on
the wave stability. Nevertheless, the excellent correspondence between
the results of the simulations in 2D and 3D suggest that for our
problem the inclusion of a third dimension would be unimportant, with regards
to wave stability.

\subsection{Implications for tidal dissipation}

In B10 and B11, we discussed the implications of the
wave breaking process for tidal dissipation in solar-type stars, and
therefore to the survival of short-period planets. What more can we
say in light of the results of this paper? 

One result is that parametric instabilities exist for waves with $A<1$. These do not
occur in an unbounded domain (the limit as $n_{p}\rightarrow \infty$),
but will be present in the RZ of a solar-type star, since this does
have an outer boundary, albeit many wavelengths from the centre of the
star. We can roughly fit Eq.~\ref{parametricgrowthrate} to the
results of our stability analysis, allowing us to give an
upper bound to the expected growth rate of the strongest instability of a tidally
excited gravity wave with $A<1$. We write
\begin{eqnarray}
\mathrm{Im}\left[\frac{\omega}{\omega_{p}} \right] \approx K\frac{A}{n_{p}},
\end{eqnarray}
and calculate a value of $K$ from the solutions to our eigenvalue
problem, where we typically find $K\approx 0.1$. In the RZ of a solar-type star, tidally excited gravity waves have
$10^{2} \lesssim n_{p}\lesssim 10^{3}$, for orbital periods in the
range $1\lesssim P \lesssim 3$ days. We can therefore calculate an upper bound on the
expected growth rate of a parametric instability in a real star from taking $A=1$ and
$n_{p}=10^{2}$, giving $\mathrm{Im}\left[\omega/\omega_{p}\right]
\approx 10^{-3}$, so that the resulting growth time,
\begin{eqnarray}
t_{grow} = \frac{1}{\mathrm{Im}\left[\omega\right]} \approx 2.7 \; \mathrm{yrs}.
\end{eqnarray}
It is important to note that this estimate is likely to be an
approximate lower bound on $t_{grow}$, and will not be strongly affected by the
inclusion of the rest of the RZ, because the
the amplitude of the waves, and therefore the nonlinearity, is much
smaller away from the centre. (In addition, our
Boussinesq-type model is only valid where $N\propto r$, which is only
true near the centre of the star.) These calculations constrain the
effects of nonlinear wave-wave interactions in the innermost regions,
but do not take into account the rest of the RZ.

It is important to estimate the magnitude of the resulting tidal
dissipation, so that we can evaluate its role in the evolution of
short-period planets. Instead of considering the problem of continual forcing of the primary
wave by the planet, we consider initialising the primary wave and 
ask how long it takes to be attenuated (and its energy dissipated), calling this timescale
$t_{nl}$ (this is similar to the highly eccentric binary problem
discussed in KG96, which we discuss in the next section). 
The torque on the star due to the gradual attenuation of the
wave due to the combined action of these parametric instabilities at
nonlinearly damping the wave, is 
\begin{eqnarray}
\Gamma = \frac{m}{\omega}F\left(1-e^{-\alpha}\right),
\end{eqnarray}
with the attenuation factor $\alpha = t_{group}/t_{nl}$, and $F$ being
computed as outlined in B10 and B11. We define the
global group travel time $t_{group} =
2\int_{0}^{r_{b}}(1/c_{g,r})dr \approx 25$ d, from a numerical
  calculation for the waves excited by
  a planet in a one-day orbit around the current Sun. The next
  question is: what is $t_{nl}$? To calculate this accurately is a
  very difficult problem, and involves many uncertainties,
  particularly those involving
  the saturation process for these nonlinear couplings. However, we
  note that a lower bound on $t_{nl}$ can be obtained by the growth rate of
  the fastest growing parametric instability $t_{grow}$. This is because this will
  act as a bottleneck for the nonlinear cascade of energy from the
  primary wave, and so will limit the maximum decay rate of the
  primary wave. This is probably also true if we are continually
  forcing the wave. We can then estimate 
\begin{eqnarray}
\alpha \lesssim \frac{t_{group}}{t_{grow}} \approx 0.025.
\end{eqnarray}
This gives an upper bound on the torque resulting from the nonlinear
damping of the primary wave. 

The resulting tidal quality factor can be computed from the expression 
\begin{eqnarray}
\label{qfactorexpression}
 Q^{\prime}_{\star} = \frac{9}{4\Gamma}\left(\frac{m_{p}}{m_{\star}+m_{p}}\right)^{2}\frac{m_{\star}R_{\star}^{2}}{\omega_{dyn}^{2}}\left(\frac{2\pi}{P}\right)^{4},
\end{eqnarray}
where $m_{\star,p}$ are the stellar and planetary masses, $R_{\star}$
is the stellar radius and $\omega_{dyn}^{2}$ is the square of the
dynamical frequency of the star $Gm_{\star}/R_{\star}^{3}$. As before,
$P$ is the planetary orbital period. This
can be used to give a lower bound on the tidal quality factor
resulting from nonlinear damping of the primary wave in the $A<1$
regime, where we find
\begin{eqnarray}
Q^{\prime}_{\star} \gtrsim \frac{10^{5}}{1-e^{-\alpha}} \approx
  \alpha^{-1}10^{5} \approx 5 \times 10^{6},
\end{eqnarray}
in the weak damping limit. The efficiency of this process is less
than critical layer absorption by a factor $\alpha^{-1} \gg 1$. Note
that this gives a lower bound on $Q^{\prime}_{\star}$, because
$t_{nl}$ is likely to be somewhat larger than $t_{grow}$ (e.g.~KG96 take
$t_{nl}=10t_{grow}$). In addition, we have taken the most optimistic
value of $A=1$, corresponding to the waves excited by a planet with a
mass of about $3M_{J}$ (see Eq.~\ref{breakingcriterion}). The resulting
$Q^{\prime}_{\star}$ may therefore be one or several orders of
magnitude larger than this lower bound. Furthermore, it is interesting to note that
this bound may not be sensitive to the number of wavelengths in the RZ,
and therefore to the orbital period, because the dependences of
$t_{group}$ and $t_{grow}$ on $n_{p}$ cancel at
leading order.

The parametric instabilities that exist when $A<1$ are much slower
than the rapid instabilities that onset when $A>1$. The nonlinear
outcome of the $A>1$ instabilities is that the wave
breaks and forms a critical layer, which then absorbs subsequent
ingoing waves, and results in astrophysically efficient tidal
dissipation. The estimate of this section indicates that the
parametric instabilities that exist when $A<1$ are much less efficient
at dissipating energy in the tide, by
several orders of magnitude. This important result supports the
explanation outlined in B10 and B11 for the survival of short-period
planets around solar-type stars.

\subsection{Comparison with Kumar \& Goodman}

We can qualitatively compare our results with previous work by
KG96, who studied nonlinear damping of tidal oscillations
in highly eccentric solar-type binaries. They used a truncated Hamiltonian approach to study
parametric instabilities of tidally excited f and g-modes.
In their model, stellar eigenmodes are coupled together from terms
that exist at third order in displacement in the expansion of the
Lagrangian density, i.e., they adopt a weakly nonlinear approach.
They consider the evolution of a mode that has been tidally excited,
but is no longer subject to forcing, due to nonlinear coupling with a
large number of g-modes that are present in the RZ of the
star (and exist because they have already been excited by turbulent convection, for example). Their
result indicates that high order and high degree g-modes can be
parametrically excited by low order quadrupolar f and g-modes, and can
draw energy from the primary mode on a timescale that is much shorter than the radiative
damping time of the primary mode.

A direct comparison of our work with theirs is not possible
for several reasons. Firstly, in their numerical work they mainly consider a
primary f-mode coupling to many g-modes in the RZ. The f-mode
eigenfunction has its largest magnitude at the surface and decays
rapidly inwards, in contrast to the primary g-modes that we are
considering, so the coupling strengths are likely to be
different. Secondly, we only consider the nonlinear
interactions in the central regions of the star, where they are likely
to be most important for g-modes, whereas they consider these
interactions throughout the whole
star. Thirdly, our model is 2D, whereas their eigenfunctions are valid
in 3D for a spherically symmetric background. This last point, however, is probably
not important.

One important point is that they neglect the possibility of wave
breaking, which would provide an upper limit to the amplitude of a
given mode. This would prevent modes with large amplitudes from
coupling with the primary wave, and the nonlinear outcome of the
breaking (most notably critical layer formation) would significantly
modify the strength of tidal dissipation. Their results will therefore
not be valid for primary or daughter waves that satisfy a breaking
criterion, 
since weakly nonlinear theory is insufficient in this case. Indeed, the concept
of parametric instability is no longer valid if the daughters break
and cannot form standing modes. This is particularly important given
their primary application of eccentric solar-type binaries, since in
that case, the ampliudes of the waves are likely to be large enough
for wave breaking near the centre of their stars (this is estimated in the
Appendix of OL07, for example).

Keeping in mind the differences between our approach and theirs, we
now directly apply their results to our problem, 
and quantitatively compare the growth time
of parametric instabilities with those found in this paper. The
growth time in their work 
\begin{eqnarray}
t_{grow} \approx 4 \left(\frac{E_{p,0}}{10^{35} \mathrm{J}} \right)^{-1/2},
\end{eqnarray}
where $E_{p,0}$ is the initial energy in the primary wave. For the
g-modes that we consider,
\begin{eqnarray}
\nonumber
E_{p,0} &=& \int \int \int \mathcal{E} r^{2}\sin \theta drd\theta
d\phi \\
&=& F \int (1/c_{g,r}) dr = F t_{group}.
\end{eqnarray}
This can be computed to give $E_{p,0} \approx 2\times 10^{29} J$ for a
Jupiter-mass planet on a one-day orbit around the current Sun, which
has $A\approx 0.3$. This
means that $t_{grow} \approx 3$ yr when $A=1$, which happens to agree
surprisingly well with our calculation in the previous section, given the
differences in our approach. The total number of daughter modes which
simultaneously interact with the primary in their model is $\sim 10^{10}
\left(\frac{E_{p,0}}{10^{35} \mathrm{J}} \right)^{5/4} \sim 10^{2}$
for our fiducial case. We also find that there are many
growing modes for a given set of parameters in our stability analysis,
so these statements appear qualitatively consistent. 
They find that collectively, these modes absorb
most of the energy of the primary wave after a time $\sim 10
t_{grow}$ (this is equivalent to assuming $\alpha \sim 2.5 \times 10^{-3}$ in
the previous section). This predicts $Q^{\prime}_{\star} \sim 5 \times 10^{7}$ in their approach. We therefore conclude that our results are broadly consistent with
KG96.

\section{Conclusions}

In this paper we have performed a stability analysis of a standing
internal gravity wave near the centre of a solar-type star, using the 2D exact
wave solution derived in BO10. This work has relevance to the tidal
interaction between short-period planets and their solar-type host
stars, since these waves are excited at the top of the radiation zone
of such a star by the tidal forcing of the planet. The equations
governing the evolution of the perturbations to this wave were written down in spectral space using
a Galerkin spectral method, and then solved as an eigenvalue
problem. This required the imposition of an artificial impermeable
outer boundary several wavelengths from the centre of the star.

We have identified the modes that initiate the breaking
process when the wave overturns the stratification. This type of mode is
strongly localised in the convectively unstable regions of the primary
wave, and is driven by unstable entropy gradients. 
Its growth time is comparable with the primary wave period,
which is consistent with the breaking time observed in the simulations
of BO10 and B11.

We have also studied the instabilities which exist for waves
with insufficient amplitudes to overturn the stratification. We find
that these are parametric instabilities driven by (convectively stable) entropy gradients in
the primary wave. The growth rate of these modes scales inversely with
the number of wavelengths within the domain, so they become less
important for a real star than for the small container considered here. It
is estimated that their growth times in a real star would be of the
order of $3$ yr, which is much longer than the orbital period of a
short-period planet, though many such modes are
excited. Rough estimates are made that provide an upper bound on the
resulting tidal dissipation, for which we find a lower bound on the
tidal quality factor of $Q^{\prime}_{\star} \gtrsim 10^{7}$ from this
process. This is much weaker than the dissipation resulting from critical layer
absorption obtained in BO10 and B11, and so is unlikely to change the
picture outlined in B11 for the survival of short-period planets.

The results of this paper provide further support for the hypothesis
outlined in BO10 and B11 for the survival of short-period extrasolar
planets around slowly rotating solar-type main-sequence stars. Coupled with weak
dissipation of the stellar equilibrium tide by turbulent convection when the
orbital period is shorter than the convective timescale (see e.g.~\citealt{Zahn1966};
\citealt{Goldreich1977}; \citealt{GoodmanOh1997};
\citealt{Penev2011}), and the
absence of inertial wave excitation in their slowly rotating stellar
hosts (OL07), it seems
likely that short-period planets can survive against tidally induced
orbital decay if they are unable to cause the internal gravity waves
that they excite to break near the centre. This paper demonstrates
that the waves need to overturn the stratification near the centre to
obtain efficient tidal dissipation.

We discussed several differences between our problem and the stability of
a plane IGW in a uniform stratification
(e.g.~\citealt{LombardRiley1996}). We have confirmed that when the wave
is confined in a container with an outer boundary it is unstable whatever its
amplitude, in the absence of diffusion. However, the inverse
dependence of the growth rate on the number of wavelengths within the
container is quite different, and results from the finite time of
nonlinear interaction being much shorter than the group
travel time across a large container. 

We compared our results to \cite{Kumar1996}, who studied the
nonlinear damping of tidally excited oscillations in highly eccentric
binaries, and found some agreement. They predict that many ($\sim 10^{2}$)
modes collectively draw energy from the primary wave, which we have
qualitatively confirmed from our stability analysis. The growth rates of parametric
instabilities for the same problem in both of our approaches when
$A\approx1$ are very similar. They therefore predict a similar lower bound for
$Q^{\prime}_{\star}$ resulting from this process. This is promising,
given the differences in our approach. It would be interesting to extend their numerical calculations by studying the parametric
instabilities of g-modes including continual tidal forcing of the
primary wave and nonlinear
couplings involving many daughter and granddaughter modes, as well as taking into
account the amplitude limiting effects of wave breaking. Weakly
nonlinear theories such as ours and theirs are likely to be valid
when considering the initial stages of the breaking process, and in
studying whether any instabilities exist for suboverturning waves,
which were the topics of study in this paper. However, they should
not be used to determine long-term behaviour for waves which overturn
the stratification (whenever $A>1$). This means that for the
circularisation of eccentric solar-type close binary stars, it is
inappropriate to use a weakly nonlinear approach, since in that case
wave breaking is very likely to occur. Instead, the results
of BO10 and B11 must be used to obtain the correct magnitude of the
dissipation, and the resulting circularisation rate due to nonlinear
interactions between gravity waves.

It would be worthwhile to confirm the results of this paper using
2D numerical simulations with SNOOPY, such as those described in BO10. An artificial impermeable outer boundary could be
implemented in the code, and the resulting instabilities then
studied. Of particular importance is to determine the rate at which
energy is lost from the primary wave due to the parametric instabilities
for suboverturning waves that we studied in this paper (i.e.,
to numerically calculate $t_{nl}$). This would
enable a more accurate calculation of the magnitude of
$Q^{\prime}_{\star}$ and would provide a useful independent check of our results. We defer such calculations to future work.

\section*{Acknowledgments}
AJB would like to thank STFC and the Cambridge Philosophical Society for a research
studentship, as well as the referee, St\'{e}phane Mathis, for a careful
reading of the manuscript.

\appendix

\section[]{Toy model: parametric instability of primary wave}
\label{parametricinstability}

Parametric instability is a type of resonant triad interaction in
which the transfer of energy from a parent (subscript $p$) mode, with
amplitude $A_{p}$, destabilises a pair of daughter (subscript $d1,d2$) modes (which exist when
$A_{p}=0$). These can then be damped or
subject to further nonlinear interactions (to produce granddaughter
modes, and so on). The frequencies of the
modes must satisfy an approximate temporal resonance condition
$\omega_{p} \approx \omega_{d1} + \omega_{d2}$, for parametric resonance to occur.

The equations governing the temporal evolution of the mode amplitudes
take the form (e.g.~\citealt{Dziembowski1982}; \citealt{WuGoldreich2001})
\begin{eqnarray}
\dot{A}_{p} &=& \gamma_{p}A_{p} - i \omega_{p} A_{p} + i \omega_{p}
\sigma A_{d1}A_{d2}, \\
\dot{A}_{d1} &=& -\gamma_{d1}A_{d1} - i \omega_{d1} A_{d1} + i \omega_{d1}
\sigma A_{p}A^{*}_{d2}, \\
\dot{A}_{d2} &=& -\gamma_{d2}A_{d2} - i \omega_{d2} A_{d2} + i \omega_{d2}
\sigma A^{*}_{d1}A_{p}.
\end{eqnarray}
In these equations, $\gamma_{j}$ is the linear growth/damping rate of
mode $j$, and $\sigma$ is the nonlinear coupling strength for these
three modes. Here $A_{j}$ is the amplitude of mode $j$, with the
energy in that mode being proportional to $|A_{j}|^{2}$.

The coupling coefficient $\sigma$ is largest when
$\omega_{d1}\approx\omega_{d2} \equiv \omega$, and therefore $\omega \approx
\omega_{p}/2$. 
If the daughter modes have similar frequency, then we can assume that
they have similar spatial scales. Hence we can take their
damping rates to be the same, i.e., $\gamma_{d1} =
\gamma_{d2} \equiv \gamma$. To consider the initial stages
of the breaking process we take $A_{p}$ to be approximately constant
in time. In our problem the primary wave is maintained at a constant amplitude due to forcing, and is not unstable. The evolutionary equations reduce to
\begin{eqnarray}
\dot A_{d1} &=& -\gamma A_{d1} - i\omega A_{d1} + i\omega \sigma A_{p}A_{d2}^{*}, \\
\dot A_{d2} &=& -\gamma A_{d2} - i\omega A_{d2} + i\omega \sigma A_{d1}^{*}A_{p}.
\end{eqnarray}
If we take $A_{d1} \propto \exp s t$, then the growth rate is
\begin{eqnarray}
\mathrm{Re}\left[s\right] = -\frac{\gamma}{2} + \frac{1}{2}\omega\sigma |A_{p}|.
\end{eqnarray}
The growth rate is reduced if the detuning $\Delta \omega = \omega_{p}
- \omega_{d1}-\omega_{d2} \ne 0$, by changing the second term to
$(1/2)\sqrt{\omega_{d1}\omega_{d2}\sigma^{2}|A_{p}|^{2}-(\Delta\omega)^{2}}$. From this model, we expect $\gamma\ne 0$ to simply reduce the growth
rate for a given mode. In addition the growth rate
scales linearly with the amplitude of the primary (parent) mode. The
threshold amplitude for instability in this simple model is
\begin{eqnarray}
\label{thresholdparametric}
|A_{p}| \geq \frac{\gamma}{\omega \sigma},
\end{eqnarray}
which depends on the coupling strength $\sigma$.

The spatial dependence of the interaction is contained in the coupling coefficient $\sigma$, which contains an integral of the product of the three eigenfunctions. This toy model of parametric instability is useful as a
simple model to understand some of the results of
\S~\ref{parametricresults}. It is interesting to note that in this model, $\omega
\approx m/k_{m,n} \sim n_{p}^{-1}$ for $A\ll 1$, since in this limit the daughter
modes have frequencies comparable with the linear mode
frequencies. This results in a growth rate scaling inversely with $n_{p}$.

\section[]{A single IGW is in equipartition}
\label{equipartition}

An IGW with a single value of $m$ and $n$ satisfies equipartition of kinetic and potential energies, when
integrated over a multiple of half-wavelengths, as we will now
prove. If we take $f(r) = J_{m}(k_{m,n}r)e^{im\phi}$, we can rewrite Bessel's equation in the form 
\begin{eqnarray}
\frac{1}{r}\partial_{r}\left(r\partial_{r} f\right) - \frac{m^{2}}{r^{2}}f = -k_{m,n}^{2}f.
\end{eqnarray}
After multiplying by $r f$, and then integrating over radius from
$r_{1}$ to $r_{2}$, we obtain
\begin{eqnarray}
\int_{r_{1}}^{r_{2}}\left[(\partial_{r}f)^{2} +
  \frac{m^{2}}{r^{2}}f^{2}\right]r dr =\int_{r_{1}}^{r_{2}}
k_{m,n}^{2}f^{2}dr + \left[r f \partial_{r} f\right]^{r_{2}}_{r_{1}}.
\end{eqnarray}
Since 
\begin{eqnarray}
K = \pi \int_{r_{1}}^{r_{2}}\left[(\partial_{r}f)^{2} +
  \frac{m^{2}}{r^{2}}f^{2}\right]r dr,
\end{eqnarray}
is the integrated kinetic energy, and 
\begin{eqnarray}
P = \pi \int_{r_{1}}^{r_{2}}
k_{m,n}^{2}f^{2}dr,
\end{eqnarray}
is the integrated potential energy, as defined in the text, this statement is telling us that
equipartition holds if we integrate over a range
where $f$ or $\partial_{r}f$ are zero at the end points. 

\bibliography{tidbib}

\begin{thebibliography}{}

\bibitem[\protect\citeauthoryear{{Barker}}{{Barker}}{2011}]{Barker2011}
{Barker} A.~J.,  2011, MNRAS, 414, 1365

\bibitem[\protect\citeauthoryear{{Barker} \& {Ogilvie}}{{Barker} \&
  {Ogilvie}}{2009}]{Barker2009}
{Barker} A.~J.,  {Ogilvie} G.~I.,  2009, MNRAS, 395, 2268

\bibitem[\protect\citeauthoryear{{Barker} \& {Ogilvie}}{{Barker} \&
  {Ogilvie}}{2010}]{Barker2010}
{Barker} A.~J.,  {Ogilvie} G.~I.,  2010, MNRAS, 404, 1849

\bibitem[\protect\citeauthoryear{{Counselman}
  III}{{Counselman}}{1973}]{Counselman1973}
{Counselman} III C.~C.,  1973, ApJ, 180, 307

\bibitem[\protect\citeauthoryear{{Drazin}}{{Drazin}}{1977}]{Drazin1977}
{Drazin} P.~G.,  1977, Proc. R. Soc. Lond. A., 356, 411

\bibitem[\protect\citeauthoryear{{Dziembowski}}{{Dziembowski}}{1982}]{Dziembow%
ski1982}
{Dziembowski} W.,  1982, Acta Astronomica, 32, 147

\bibitem[\protect\citeauthoryear{{Fritts}, {Wang}, {Werne}, {Lund} \&
  {Wan}}{{Fritts} et~al.}{2009}]{Fritts2009}
{Fritts} D.~C.,  {Wang} L.,  {Werne} J.,  {Lund} T.,    {Wan} K.,  2009, J.
  Atmos. Sci., 66, 1126

\bibitem[\protect\citeauthoryear{{Goldreich} \& {Nicholson}}{{Goldreich} \&
  {Nicholson}}{1977}]{Goldreich1977}
{Goldreich} P.,  {Nicholson} P.~D.,  1977, Icarus, 30, 301

\bibitem[\protect\citeauthoryear{{Goldreich} \& {Soter}}{{Goldreich} \&
  {Soter}}{1966}]{GoldSot1966}
{Goldreich} P.,  {Soter} S.,  1966, Icarus, 5, 375

\bibitem[\protect\citeauthoryear{{Goodman} \& {Dickson}}{{Goodman} \&
  {Dickson}}{1998}]{GoodmanDickson1998}
{Goodman} J.,  {Dickson} E.~S.,  1998, ApJ, 507, 938

\bibitem[\protect\citeauthoryear{{Goodman} \& {Oh}}{{Goodman} \&
  {Oh}}{1997}]{GoodmanOh1997}
{Goodman} J.,  {Oh} S.~P.,  1997, ApJ, 486, 403

\bibitem[\protect\citeauthoryear{Hasselmann}{Hasselmann}{1967}]{Hasselman1967}
Hasselmann K.,  1967, J. Fluid Mech., 30, 737

\bibitem[\protect\citeauthoryear{{Hut}}{{Hut}}{1980}]{Hut1980}
{Hut} P.,  1980, A\&A, 92, 167

\bibitem[\protect\citeauthoryear{{Klostermeyer}}{{Klostermeyer}}{1982}]{Kloste%
rmeyer1982}
{Klostermeyer} J.,  1982, J. Fluid Mech., 119, 367

\bibitem[\protect\citeauthoryear{{Koudella} \& {Staquet}}{{Koudella} \&
  {Staquet}}{2006}]{Koudella2006}
{Koudella} C.~R.,  {Staquet} C.,  2006, J. Fluid Mech., 548, 165

\bibitem[\protect\citeauthoryear{{Kumar} \& {Goodman}}{{Kumar} \&
  {Goodman}}{1996}]{Kumar1996}
{Kumar} P.,  {Goodman} J.,  1996, ApJ, 466, 946

\bibitem[\protect\citeauthoryear{{Landau} \& {Lifshitz}}{{Landau} \&
  {Lifshitz}}{1969}]{Landau1969}
{Landau} L.~D.,  {Lifshitz} E.~M.,  1969, {Mechanics}

\bibitem[\protect\citeauthoryear{{Lombard} \& {Riley}}{{Lombard} \&
  {Riley}}{1996}]{LombardRiley1996}
{Lombard} P.~N.,  {Riley} J.~J.,  1996, Physics of Fluids, 8, 3271

\bibitem[\protect\citeauthoryear{{McEwan} \& {Robinson}}{{McEwan} \&
  {Robinson}}{1975}]{McEwanRobinson1975}
{McEwan} A.~D.,  {Robinson} R.~M.,  1975, J. Fluid Mech., 67, 667

\bibitem[\protect\citeauthoryear{{Meid}}{{Meid}}{1976}]{Meid1976}
{Meid} R.~P.,  1976, J. Fluid Mech., 78, 763

\bibitem[\protect\citeauthoryear{{Ogilvie} \& {Lin}}{{Ogilvie} \&
  {Lin}}{2007}]{Gio2007}
{Ogilvie} G.~I.,  {Lin} D.~N.~C.,  2007, ApJ, 661, 1180

\bibitem[\protect\citeauthoryear{{Penev} \& {Sasselov}}{{Penev} \&
  {Sasselov}}{2011}]{Penev2011}
{Penev} K.,  {Sasselov} D.,  2011, ApJ, 731, 67

\bibitem[\protect\citeauthoryear{{Sonmor} \& {Klaassen}}{{Sonmor} \&
  {Klaassen}}{1997}]{SonmorKlaassen1997}
{Sonmor} L.~J.,  {Klaassen} G.~P.,  1997, J. Atmos. Sci., 54, 2655

\bibitem[\protect\citeauthoryear{{Staquet} \& {Sommeria}}{{Staquet} \&
  {Sommeria}}{2002}]{Staquet2002}
{Staquet} C.,  {Sommeria} J.,  2002, Ann. Rev. Fluid Mech., 34, 559

\bibitem[\protect\citeauthoryear{{Terquem}, {Papaloizou}, {Nelson} \&
  {Lin}}{{Terquem} et~al.}{1998}]{Terquem1998}
{Terquem} C.,  {Papaloizou} J.~C.~B.,  {Nelson} R.~P.,    {Lin} D.~N.~C.,
  1998, ApJ, 502, 788

\bibitem[\protect\citeauthoryear{{Wu} \& {Goldreich}}{{Wu} \&
  {Goldreich}}{2001}]{WuGoldreich2001}
{Wu} Y.,  {Goldreich} P.,  2001, ApJ, 546, 469

\bibitem[\protect\citeauthoryear{{Zahn}}{{Zahn}}{1966}]{Zahn1966}
{Zahn} J.~P.,  1966, Annales d'Astrophysique, 29, 489

\end{thebibliography}

\end{document}